# Modular AI agents for transportation surveys and interviews: Advancing engagement, transparency, and cost efficiency

Jiangbo Yu[a, *], Jinhua Zhao[b], Luis Miranda-Moreno[a], Matthew Korp[c]

[a] *Department of Civil Engineering, McGill University, Montreal, H3A 0C3, Canada*

[b] *Department of Urban Studies and Planning, Massachusetts Institute of Technology, Cambridge, 02139, USA*

[c] *Department of Mechanical Engineering, McGill University, Montreal, H3A 0C3, Canada*

**Corresponding author.

E-mail: jiangbo.yu@mcgill.ca

**Abstract:** Surveys and interviews—structured, semi-structured, or unstructured—are widely used for collecting insights on emerging or hypothetical scenarios. Traditional human-led methods often face challenges related to cost, scalability, and consistency. For example, distributed questionnaires lack the ability to provide real-time guidance and request immediate clarifications. Recently, various domains have begun to explore the use of conversational agents (chatbots) powered by generative artificial intelligence (AI) technologies. However, considering decisions in transportation investments and policies often carry significant public and environmental stakes, surveys and interviews face unique challenges in integrating AI agents, underscoring the need for a rigorous, resource-efficient approach that enhances participant engagement and ensures privacy. This paper addresses this gap by introducing a modular approach and its resulting parameterized process for designing AI agents. We detail the system architecture, integrating engineered prompts, specialized knowledge bases, and customizable, goal-oriented conversational logic. We demonstrate the adaptability, generalizability, and efficacy of our modular approach through three empirical studies: (1) travel preference surveys, highlighting conditional logic and multimodal (voice, text, and image generation) capabilities; (2) public opinion elicitation on a newly constructed, novel infrastructure project, showcasing question customization and multilingual (English and French) capabilities; and (3) expert consultation about the impact of technologies on future transportation systems, highlighting real-time, clarification request capabilities for open-ended questions, resilience in handling erratic inputs, and efficient transcript postprocessing. The results suggest that the AI agent increases completion rates and response quality. Furthermore, the modular approach demonstrates controllability, flexibility, and robustness while addressing key ethical, privacy, security, and token consumption concerns. We believe this work lays the foundation for next-generation surveys and interviews in transportation research.

## 1. Introduction

Surveys and interviews (S&Is), whether structured, semi-structured, or unstructured, are indispensable tools for gathering insights in diverse fields, such as household surveys, public opinion elicitation, customer service, and beyond. Surveys typically involve the distribution of standardized questionnaires to collect quantitative or qualitative data from a broad population, aiming for scalability and replicability. In contrast, interviews, although they can overlap with surveys, usually refer to more conversational procedures, providing an opportunity for real-time interaction between the interviewer and the respondent. This interaction allows for open-ended responses, deeper exploration of topics, and the ability to clarify or follow up on answers, making interviews particularly valuable for gathering rich, high-quality data. While surveys focus on breadth, interviews tend to focus on depth, providing flexibility to explore individual preferences and judgments. Certain types of surveys, such as those that measure physical attributes (e.g., parking spaces and sizes of residential, commercial, and industrial zones), conduct traffic counts, process smart card data, or analyze cellular data, which do not require intercepting or actively engaging human participants, fall outside the scope of the present paper.

S&Is offer researchers a tool for measuring individual and collective perceptions, preferences, and decision-making processes. Despite their proven utility, traditional data collection techniques encounter several challenges. The resource-intensive nature of human-led surveys often leads to high costs and time inefficiencies, while their static design struggles to scale across large populations or adapt to dynamic contexts. Survey interviewers can negatively affect survey data by introducing variance and bias into estimates (Kühne, 2023) and are exposed to various risks (e.g., disease transmission, exposure to violence). Although more scalable, traditional distributed questionnaires lack the flexibility for real-time clarifications and interactive follow-ups, which are crucial for gathering nuanced responses.

Artificial Intelligence (AI) agents, sometimes referred to as conversational agents, chatbots. or digital assistants, have emerged as a viable alternative to address these limitations. Leveraging advances in natural language processing (NLP), generative algorithms, deep learning, and reinforcement learning, AI agents can engage users in a more interactive and responsive manner, simulating human-like conversations. Studies outside the transportation domains have demonstrated that compared with conventional S&I methods, AI agents might drive higher engagement rates and elicit more detailed responses, particularly in scenarios involving open-ended questions and dynamic feedback loops (Caldarini et al., 2022; Xiao et al., 2020). Their ability to maintain consistency in tone and adapt to user input in real time makes them particularly suitable for environments where personalized and context-sensitive interactions are valued (Ngai et al., 2021).

However, existing studies focus primarily on serving customers' intent (as in customer services and support), whereas significantly fewer studies explore the use of AI agents for various purposes in S&Is, which need to be, at least partially, driven by the AI agent's intent (rather than the human user's intent). When a participant digresses to a different topic when answering a question about their daily commute experience, an agent should recognize that and try to redirect the conversation to the original topic (instead of following the participant's digressed topic). In addition to the insufficient research on using AI agents for S&Is, current implementations often suffer from limitations in scalability, adaptability, domain-specific application, and privacy concerns. Most existing AI agent systems are designed for singular, narrowly defined tasks, which limits their reusability across different contexts and research domains. For example, while some chatbots excel in healthcare settings by efficiently triaging patient inquiries, as demonstrated by Lee et al. (2022), others focus solely on customer service interactions in e-commerce without showing their ability to adapt their logic to different industries or use cases (Adam et al., 2021). Indeed, the transferability of the techniques or logic flows for improving customer compliance or experience to S&Is is unknown. Although recent advancements in generative artificial intelligence (generative AI), particularly the integration of large language models (LLMs) have significantly enhanced chatbot capabilities, enabling them to understand and generate more complex and contextually relevant responses, the absence of a generalizable, privacy-preserving approach to leverage these advancements has hindered their application and cross-domain scalability in S&I.

This paper aims to fill this gap by introducing a modular approach to designing AI agents powered by LLMs tailored for eliciting opinions and judgments from mobility service users and transportation professionals. Unlike prior implementations, which are often developed for specific S&Is use cases, our proposed framework emphasizes flexibility, reusability, and adaptability for a wide range of purposes in transportation S&Is, including stated or revealed travel preference surveys (e.g., about destinations, modes, routes, and telework), public opinion and perception elicitation (e.g., about projects, policies, and technology), and professional and expert consultation (e.g., about current industry practice and the impacts of technology, policy, and disruptive events). We present a holistic design that integrates engineered prompts, specialized knowledge bases, and customizable, goal-oriented conversational logic, thereby enabling the agent to handle variations in questioning and dynamic follow-ups efficiently for various transportation research purposes. The modular approach allows an explicit description of how the internal modular logic components within each question round (a conversational flow regarding a specific question or a topic) interact and is conditioned on prior question rounds and the overall objective. Importantly, when the design of a survey or interview process is inappropriate, the modular approach should facilitate the exposure of mistakes or unsuitability. We also propose that one of the modules should be to store key information from a user response to a variable instead of saving the entire response to reduce AI/LLM token consumption, which will remain an important constraint for large-scale deployment of AI agents in the foreseeable future. In addition, we address key ethical, privacy, and security considerations that arise when deploying AI-powered conversational systems in sensitive contexts. As different S&I purposes have different levels of feature desirability, we summarize them in Table 1.

We demonstrate the modular framework's versatility and effectiveness in three concrete field studies: (1) resident preference surveys, (2) public opinion collection, and (3) expert/professional opinion elicitation. Each of these studies showcases the unique capabilities of an AI agent, such as multilingual interaction, multimodal support, and adaptive questioning strategies. By presenting this framework, we seek to provide a robust foundation for future research and development in the field of AI agents for S&I, extending their applicability beyond singular use cases to a broader range of data collection and interactive applications. We believe that the framework also helps set the stage for a new generation of intelligent, adaptable, and scalable conversational systems for human-AI collaboration where a task or a project is driven by a combination of human/customer intent and AI intent.

**Table 1** Level of desirability for key features in the three most common types of S&Is in transportation research.

| Desirable feature | Travel preference | Public opinion & perception | Professional consultation |
|---|---|---|---|
| Open-ended questions | ++ | +++ | +++ |
| Customizable questions | +++ | +++ | ++ |
| Tailored follow-up questions | ++ | +++ | +++ |
| Scalable | +++ | +++ | + |
| Multilingual | +++ | +++ | + |
| Multimodal | ++ | +++ | + |
| Domain knowledge | + | + | +++ |
| Identity preserving | +++ | +++ | ++ |
| Automated postprocessing | +++ | +++ | + |

Note: +: Not a bottleneck (yet); ++: Moderate; +++: Strongly needed.

## 2. Related work

The evolution of AI agents has garnered significant attention in recent years, driven by advancements in NLP, machine learning (ML), high-performance computing, and high-speed telecommunication. As these interconnected technologies continue to improve, the applications of AI agents have expanded rapidly across diverse domains, including healthcare, education, and customer service. These developments underscore the potential of AI agents to transform how information is obtained and delivered. This section first reviews the transportation literature related to S&Is. Then we synthesize key studies on AI applications, with a focus on their conversational capabilities, and their emerging use in S&I to date.

### 2.1. Surveys and interviews in transportation research

S&Is have long been foundational methods in transportation research, offering valuable insights into travel behavior, public opinion, and expert judgment. S&Is are frequently employed to collect quantitative data on individuals' revealed and stated travel preferences regarding destination, mode, telecommuting, value-of-time, and departure time, along with their backgrounds (e.g., socioeconomic, demographic, and physiological) (Campbell et al., 2016; Loong et al., 2017; Zahabi et al., 2015). Many common yet essential revealed preference surveys, such as household and origin-destination travel surveys, are often costly and may take many years to update officially, resulting in policy and investment decisions based on projections derived from severely outdated information. Stated preference surveys often incorporate discrete choice experiments to simulate decision-making environments—whether based on memory recall or described "what if" scenarios—where respondents are presented with controlled alternatives and asked to express their preferences (Kontou et al., 2017; Yu and Jayakrishnan, 2018). This method enables researchers to estimate the trade-offs travelers are willing to make between different attributes, such as cost versus convenience, time versus environmental impact, or purchasing versus renting a car (Shen et al., 2016). Traditional methods, such as household travel surveys, workplace interviews, and on-board transit surveys, have evolved significantly with technological advancements, offering greater precision and efficiency through web-based and mobile app-based surveys (Wang et al., 2021). These innovations help address common challenges such as nonresponse bias and recall accuracy, which often affect self-reported travel data. While location-based service data are increasingly available for real-world travel observations, surveys also remain a crucial benchmark and reference tool in non-hypothetical scenarios (e.g., which mode did you use for commuting yesterday?) (Toole et al., 2015). Web-based surveys were especially vital for understanding shifting travel needs during the COVID-19 pandemic in 2020 (Irawan et al., 2022). The data collected from these surveys are instrumental in informing demand models, such as random utility-based discrete choice models or emerging machine learning approaches (Wang and Ross, 2018). These results can be analyzed at the discrete, individual (microscopic) level (Ding et al., 2018) or aggregate level through spatialtemporal analytics and enhancement via other data sources, such as location-based service data, virtual/augmented reality data (Farooq and Cherchi, 2024), driving simulators (Harb et al., 2018), and others. However, issues associated with high administrative costs, framing and experimenter effects, and privacy preservation remain (Barthelmes et al., 2023; LaMondia and Fisher, 2024).

In addition to understanding travel behaviors, S&Is are frequently used to gauge public perceptions of emerging transportation technologies, infrastructure projects, and urban policies. For instance, S&Is have been instrumental in assessing public attitudes toward ridesharing services (Krueger et al., 2016), autonomous driving technologies (Bansal and Kockelman, 2018; Harper et al., 2016; Kyriakidis et al., 2015; Madigan et al., 2017), and the adoption of electric vehicles and their associated charging infrastructure (Greene et al., 2020; Helveston et al., 2015). They also provide valuable insights into public sentiment toward hypothetical disruptive events, such as earthquakes or epidemics (Lindell et al., 2020), and policies under consideration, such as congestion pricing (Hess and Börjesson, 2019). Public consultations in the form of online posts that allow direct commenting or email responses, often seen as a form of S&Is, are widely employed by government agencies to inform policy and investment decisions and elicit feedback from the public. However, large-scale consultations are sometimes criticized as tokenistic, particularly when agencies lack the capacity to meaningfully address or respond to the substantial volume of feedback received. For example, initiatives like *Consulting with Canadians* have been criticized as superficial due to the limited human resources available to public agencies, often leading to consultations being conducted without the necessity of responding to or engaging in meaningful dialog with participants (al Attar and Clouthier, 2015; Longo, 2017). Despite these challenges and limitations, S&Is remain indispensable in transportation research.

S&Is are also frequently employed to elicit domain expertise, in-depth insights, and professional judgments from specific stakeholder groups, such as transportation engineers, urban planners, policymakers, community representatives, transit operators, technology company executives, and academic researchers (Aba and Esztergár-Kiss, 2024; Adhikari et al., 2020; Karolemeas et al., 2021; Kester et al., 2018; Nævestad et al., 2015; Pilko et al., 2017; Whittle et al., 2019; Zhang et al., 2021). These S&Is can also be part of a participatory decision-making process related to transportation investments or policies (Ghorbanzadeh et al., 2018; Yu and McKinley, 2024). These S&Is provide critical qualitative data that extend beyond public opinion by drawing on the specialized knowledge and experiences of individuals who influence the development and implementation of transportation systems and policies. Interviews, conducted in person or virtually, are particularly effective in this context because of their flexible structure, which combines open-ended questions with a semi-structured interview guide. This approach enables interviewers to explore complex issues in depth while maintaining a focused discussion on key topics. Semi-structured interviews are especially valuable in capturing the nuances of expert judgment; revealing motivations; personal experiences; and professional attitudes toward transportation-related challenges, policies, and technologies while maintaining some standardization and control over the interview process. Effective interviewing in this context requires a degree of expertise from the interviewer, who must be capable of asking follow-up questions and seeking clarification when necessary. Another commonly employed method for expert elicitation is the Delphi method, which involves multiple rounds of surveys to achieve consensus among experts on complex issues (Darcy and Burke, 2018). Compared with travel surveys and public perception studies, expert opinion collection often involves smaller sample sizes due to the specialized nature of the participants. The data collected tend to be more detailed and nuanced, typically expressed through natural language responses, which require extensive postprocessing. Unlike quantitative data, which can be directly input into statistical models, expert interviews generally require qualitative techniques such as content analysis, coding, or thematic analysis to identify recurring patterns and extract actionable insights. This qualitative approach is critical for gaining a deeper understanding of complex issues, particularly when assessing the feasibility and impact of transportation innovations or policy interventions. Expert elicitation through S&Is plays a crucial role in evaluating the practical implications of emerging technologies, such as autonomous vehicles, electric mobility, and smart infrastructure, as well as in identifying regulatory challenges and potential barriers to implementation. By incorporating the expertise of professionals directly involved in these fields, researchers can gain a more comprehensive understanding of transportation systems and use these insights to inform policy and investment decisions. This expert judgment is often combined with quantitative models to offer a more holistic view, enabling better decision-making in both research and policy contexts. However, in addition to the challenges associated with travel preferences and public perceptions of S&Is, challenges such as interview staff training, professional/expert identification, and transcript postprocessing are particularly impeding (Solarino and Aguinis, 2021).

### 2.2. Applications of AI agents

Preliminary attempts to use AI agents have been made in diverse fields, each highlighting their potential to enhance scalability, real-time interactivity, and automated processes, although challenges remain in ensuring user trust, satisfaction, and seamless integration into more personalized or sensitive interactions (Solarino and Aguinis, 2021). In education, conversational agents have been integrated into classroom settings to support students' learning experiences. Chen et al. (2023) explored the use of AI-driven student assistants in business education and find that chatbots help students learn basic content and provide insights into student needs. Similarly, Belda-Medina and Calvo-Ferrer (2022) examined the role of chatbots in language learning and discovered that students had positive perceptions of ease of use and ease of interaction, although their behavioral intention to use them remained moderate. In the customer service and e-commerce sectors, chatbots have significantly transformed the customer experience. Ngai et al. (2021) proposed an intelligent, knowledge-based conversational agent system for customer service in e-commerce. Hsu and Lin (2023) further examined user satisfaction and loyalty in customer service chatbots and revealed that both the core service quality and the conversational quality of AI chatbots significantly influence customer satisfaction and loyalty. Conversational agents are also gaining traction in social media. Brandtzaeg et al. (2022) investigated how users form friendships with social chatbots such as Replika, identifying parallels between human-AI and human-human friendships, with chatbots offering personalized interactions that fulfill users' needs. Moreover, in healthcare, chatbots have been utilized to support both physical and mental health care. Lee et al. (2022) demonstrated how chatbots can help manage hypertension in clinical settings, reducing clinician workloads by effectively triaging patient messages. Moilanen et al. (2023) explored the role of chatbots in supporting mental health self-care and discovered that users’ trust was contingent upon the perceived security and integrity of chatbot systems. De Freitas et al. (2024) highlighted that LLM-powered chatbots, while engaging in mental health-related conversations, often provide unhelpful or risky responses, posing significant safety risks for vulnerable users. These findings underscore the need for cautious deployment and regulation of AI companions in mental health contexts. Benaddi et al. (2024) explored the broad application of chatbots in the tourism sector, noting that while chatbots have been widely adopted for customer interactions, academic research on their technological evolution and impact remains limited. Their systematic review offers a comprehensive classification of chatbot architectures and assesses their key applications in tourism, where they have been integrated to enhance responsiveness and user satisfaction.

The deployment of (conversational) AI agents in the transportation sector is expanding, with growing applications in autonomous vehicles, participatory planning, and decision support. Kim et al. (2024) examined the use of human-like voice agents in self-driving cars and find that increased anthropomorphism can enhance intimacy and perceived human-like qualities but does not significantly affect trust or safety perceptions. This complexity underscores the importance of balancing emotional engagement with functional efficiency in voice agent design for autonomous vehicles. Liao et al. (2024) introduced an encoder-decoder framework augmented by LLMs to address visual grounding in autonomous vehicles. Yu and McKinley (2024) introduced a synthetic participatory approach to facilitate ideation and strategic investment in future mobility systems. By employing conversational agents to represent diverse human stakeholders to converse with one another and with real human stakeholders, they demonstrated the potential for enhancing inclusivity and strategic planning in complex, multi-stakeholder, and multi-objective decision-making problems.

LLMs have significantly enhanced the conversational abilities of AI agents, enabling more natural, complex, and human-like interactions (Myers et al., 2024). Traditional agents, which rely on rule-based systems or narrow machine learning techniques, often struggle to understand nuanced inputs or maintain fluid conversational flows. LLMs such as ChatGPT and DeepSeek have transformed this landscape by offering advanced capabilities in NLP, context retention, and logical inference to generate coherent, context-aware responses based on user inputs. Many of these models have demonstrated remarkable proficiency in generating coherent, context-aware responses on the basis of user inputs. The core module behind many LLMs is the self-attention mechanism in transformers, which serves as the fundamental building block for language modeling tasks. Reinforcement learning from human feedback (RLHF) is another crucial aspect of LLMs and involves fine-tuning the model via human-generated responses as rewards/penalties. This process allows the model to learn from its mistakes and improve its performance over time (Meshkin et al. 2024). Mechanisms such as diffusion, generative adversarial networks, and allowing external packages or services further enhance the multimodal ability of AI agents such as generating images, voices, and videos and using external references or services (Sun et al., 2024).

One of the most impactful features of LLMs in AI agent development is their ability to process vast amounts of unstructured data, allowing chatbots to generate highly accurate, human-like responses. Unlike earlier systems, LLM-powered agents excel at handling open-ended queries, understanding complex grammar and syntax, and maintaining consistency over long, multiturn conversations. Agent developers can select online (cloud-based) LLMs and on-premise (locally deployed) LLMs to suit the security, privacy, and other needs of enterprises (Meshkin et al., 2024). Retrieval Augmented Generation (RAG) is a hybrid architecture that combines retrieval mechanisms with the generative capabilities of LLMs to enhance the domain knowledge and expertise of a conversational agent (Alsafari et al., 2024). Additionally, LLMs enhanced by mechanisms such as diffusion models and generative adversarial networks (GANs) enable high-performance multimodal capabilities, allowing agents to process and generate diverse outputs such as voices and images. This multimodality is particularly valuable in fields such as urban planning or public opinion research, where diverse input types can enrich data collection and user engagement. Furthermore, LLM-powered agents can support multiple languages (multilingual capability), which is crucial when S&I participants are highly heterogenous and prefer to communicate in different languages (Miah et al., 2024). Using LLMs for role-playing has recently drawn attention, as agents can simulate various personas or stakeholders (Gao et al., 2024; Shanahan et al., 2023; Yu and McKinley, 2024). This usage not only has broad scope of AI agent use cases but also offers a unique perspective for examining their capabilities and societal implications.

### 2.3. AI agents for Surveys and interviews

The use of AI agents for conducting S&Is outside of transportation domains has gained traction in recent years because of their scalable and interactive nature, which allows for immediate feedback and simultaneous engagement with a large number of participants with real-time updates. Xiao et al. (2020) compared chatbot-based surveys with traditional online survey tools, focusing on open-ended questions. They reported that chatbots significantly improved participation rates and response quality, as measured against Gricean Maxims of informativeness, relevance, specificity, and clarity. However, this study primarily examined the contrast between chatbots and conventional tools without detailing how the chatbot managed the logic and information flow during S&I. For example, it remains unclear how the system determines response sufficiency and when to transition to the next question. Moreover, the case studies were limited to the entertainment sector (preferences for video games and movies), leaving questions about the chatbot's applicability in more complex domains, such as infrastructure investment or public policy, unanswered. Rhim et al. (2022) explored how humanization techniques in chatbot design impact respondents' experiences, particularly in terms of perceptions, interaction quality, and data output. Meanwhile, Zhang and Konka (2024) used a chatbot to assess household resilience during winter storms, whereas Jiang et al. (2023) introduced "CommunityBots", a multi-agent platform that aims to enhance engagement and data quality through efficient topic switching and user intent management. Dortheimer et al. (2024) tested a proof-of-concept chatbot for engaging participants on a hypothetical architecture project, with participants preferring the chatbot to traditional surveys. Yuan et al., 2024) examined chatbots in personality assessments and reported that both AI-inferred and traditional methods capture different personality aspects, suggesting that both approaches have value for real-world applications.

Despite these advancements, the aforementioned studies in this subsection primarily rely on illustrative examples without disclosing detailed chatbot architectures or information management procedures. This lack of empirical decision problems and transparency in system configurations complicates rather than simplifies the evaluation of how AI agents conduct S&Is and the conditions under which reported outcomes were derived. Indeed, Zarouali et al. (2024) reported that traditional web surveys outperformed chatbots in terms of data quality, perceived enjoyment, usefulness, and security, raising questions about the efficacy of chatbots as valid survey alternatives. On the other hand, Sachdeva et al. (2024) reported mixed results, depending on the user type and survey design. Furthermore, as Jin et al. (2024) noted, even when the chatbot architecture and procedures remain constant, the prompting strategies employed can significantly influence survey outcomes. Without a transparent, explicit description of conversational systems, disentangling different factors and addressing the mixed findings remain challenging.

The use of AI agents in S&Is holds significant potential, but challenges persist due to the lack of explicit system architectures and clear procedural frameworks for handling single- and multi-topic questions. These challenges are particularly relevant for the development and adoption of AI agents in transportation S&Is, where the complexity of topics and the need for reliable data are critical. The absence of structured frameworks not only increases development costs but also risks uncertain survey quality, particularly in high-stakes contexts like infrastructure planning and urban policy. Deploying these agents in sensitive areas, such as political analysis or public investment decisions, amplifies privacy and ethical concerns, especially when relying on opaque commercial LLMs. The use of transformer-based LLMs in commercial applications has raised issues of legal accountability, necessitating robust regulatory frameworks (Migliorini, 2024). Ethical guidelines emphasizing transparency, accountability, and fairness are critical to prevent misuse (Ghandour et al., 2024). Furthermore, current LLMs face limitations in complex pragmatic tasks, such as emotion recognition and sentiment analysis, despite their proficiency in generating fluent text (Kocoń et al., 2023). These limitations underscore the need for real-time bias detection and correction mechanisms to ensure fairness and accuracy in S&I tasks. A standardized approach—featuring transparent architectures, well-defined procedures, and ethical safeguards—is essential to address these challenges and advance the adoption of AI agents in transportation S&Is.

## 3. System and process

In this section, we propose a system architecture for an AI agent conduct S&Is. We elaborate on the key modules (components) outlining their roles within the overall framework and describe the information flow that governs the agent's operations during S&Is. Additionally, we describe design principles that ensure effective, consistent, secure, and token-efficient, optimizing data collection while maintaining respondent engagement and privacy. Furthermore, we introduce a formal mathematical description of the system's operational dynamics, greatly facilitated through the modular approach. To the best of our knowledge, this represents the first rigorous attempt to systematically document a AI agent for S&Is and systematically characterize the interaction patterns of the agent, providing a foundation for future research and development in this domain.

### 3.1. Proposed system architecture and configuration

The proposed architecture of AI agents for S&Is consists of five main modules: LLMs, a user interface (UI) (or a set of UIs for multiple types of devices), engineered prompts, specialized knowledge bases, session variable storage, and customized conversational logic ranging from simple procedures without referring to conversational history to highly brunched or cross-referenced procedures. The connections of these modules are illustrated in Fig. 1. To protect privacy, user utterance is first examined by an on-premise, open-sourced (and perhaps less powerful) LLM. If the utterance is deemed to have no identity-exposing or risk-imposing information, the utterance can be sent to an online (and perhaps more powerful) LLM for further processing. Depending on the capability of LLMs and the specific application context, external resources, such as specialized software and cloud computing services, can also be called by the process manager or the LLMs to further enhance the functionality of the overall conversational system. Engineered prompts or prompt templates are carefully designed input queries that guide the agent's interaction with users. These prompts or templates of prompts are constructed to ensure clarity, reduce ambiguity, and minimize biases. The prompts are structured to facilitate different types of responses, including open-ended questions, multiple-choice options, and follow-up questions for clarification. The agent utilizes specialized knowledge bases that contain domain-specific information relevant to the survey or interview context. These knowledge bases should be curated and timely updated to ensure that the agent provides accurate and current information. The knowledge bases are integrated with the agent enabled by advanced information retrieval techniques such as RAG, which allows the agent to access and query relevant data from knowledge bases (rather than from user responses or online searches).

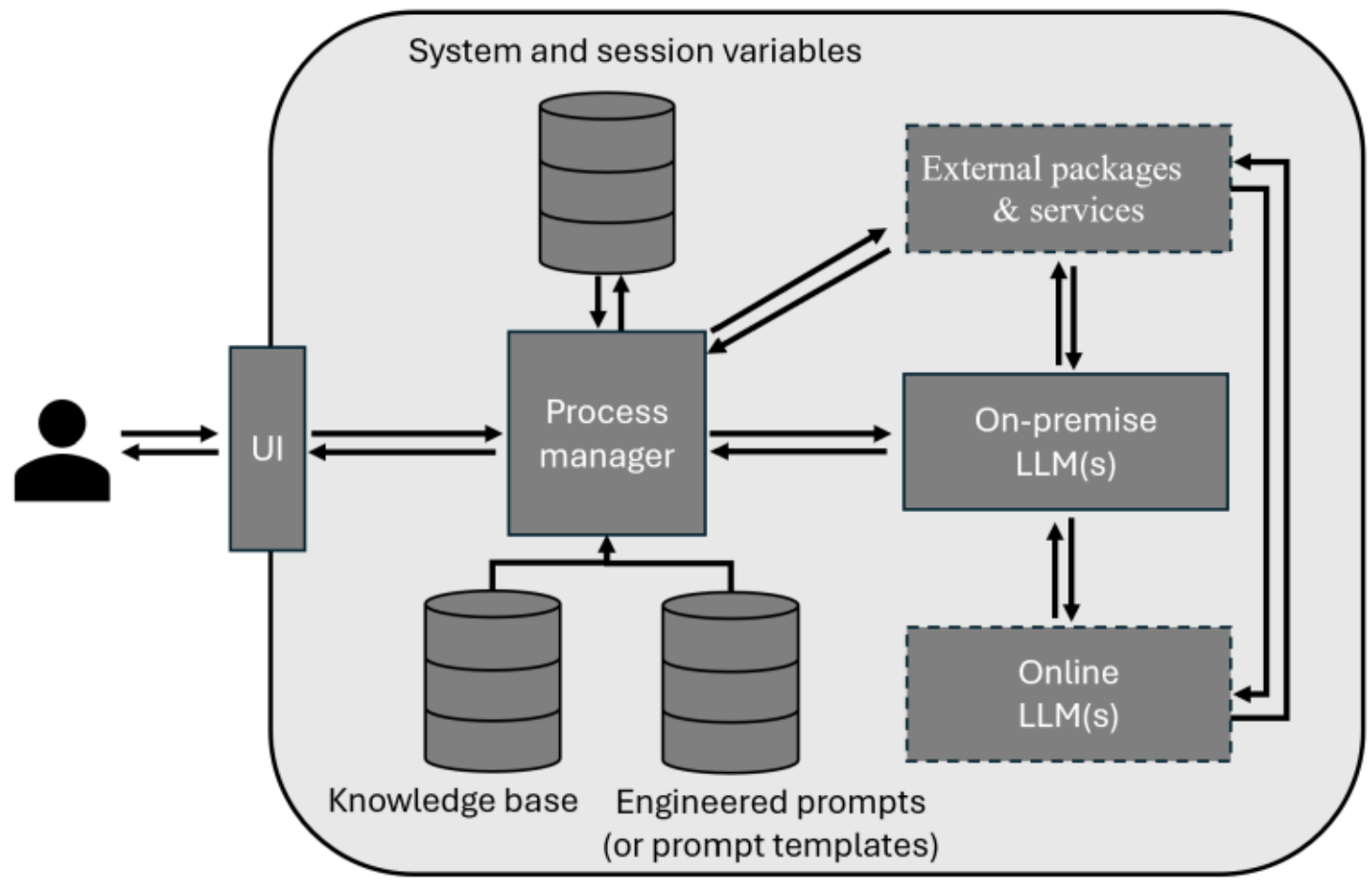

**Fig. 1** Modules and their interactions in an AI agent for surveys and interviews.

### 3.2. Parameterized process

The system modularization enables a parameterized description of how an AI agent operates during S&I, which significantly facilitates the systematic testing and documentation of an AI agent. Customized conversational flows define the sequence and structure of the interaction between an agent and a participant. These flows are designed to mimic natural conversation patterns while ensuring that the necessary information is elicited systematically with a desirable degree of replicability. The conversational flows are dynamic and can adapt on the basis of user responses, enabling the agent to ask follow-up questions, provide clarifications, and probe deeper into specific topics as needed. This process can be seen as an iterative process in which the agent attempts to elicit information (the goal) through a question and determines whether the response sufficiently answers the question (reaches the goal). If not, the agent asks for clarification and determines if the new response sufficiently addresses the question. If the new response (in addition to the previous response) is considered sufficient, the agent ends the session and proceeds with a new session (i.e., asks the next question). This next question might be predetermined or generated on the fly on the basis of the dialog thus far and the goal. For each question, the process is illustrated in Fig. 2. In structured S&Is, the agent immediately proceeds to the next (pre-determined) question and does not ask follow-up questions, which can be seen as having a 100% chance to consider the utterance sufficiently addressing the question in Fig. 2. In semi-structured S&Is, the agent might instead ask for clarifications. In unstructured interviews, the agent comes up with a question based on the dialog thus far and determines what the next question should be.

To describe the process more specifically and rigorously, we use a three-question (three-session) semi-structured interview as a concrete example. The process starts with a pre-constructed system prompt (or prompt template) $\rho_0 \in P$, system setting vector $\boldsymbol{\theta}_0 \in \Theta$, and a set of knowledge bases $\{\omega_0, \omega_1, \ldots, \omega_N\} \in \Omega$. $\rho_0$ is typically a string containing information such as "you are conducting an interview about [a certain topic]." The elements of the parameter vector $\boldsymbol{\theta}_0$ might contain the LLM version, temperature (the randomness or stochasticity of the LLM outputs), maximum allowable input and output characters, etc. $u_1 \in \mathcal{U}^q$ is the prompt for initializing the first interview question (i.e., $u_i$ and $i = 1$). To simplify the illustration, we do not consider $\omega$. $\Theta$, $\mathrm{P}$, and $\mathcal{U}$ are the parameter space, the *system* prompt space, and the prompt space, respectively. Together, $\rho_0$ and $\boldsymbol{\theta}_0$ form the initial system setting.

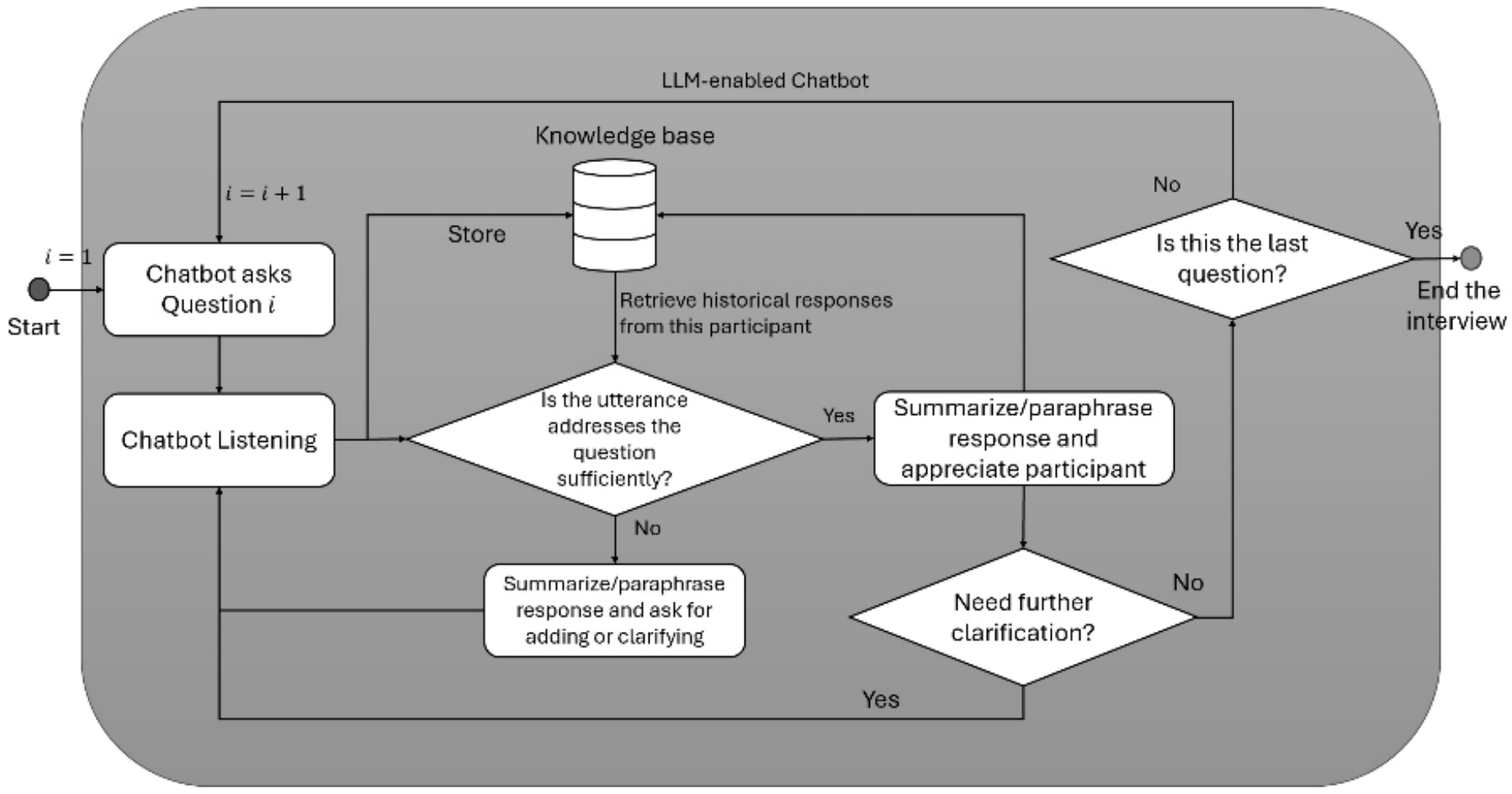

**Fig. 2** Example workflow for conducting semi-structured interview, enabled by the coordination of different system components.

Through a system setting in the space $\Theta \times \mathrm{P}$, we can use an LLM to realize a (stochastic) function $\mathcal{LM}^q: \mathcal{U}^q \to \mathcal{Q}$, where $\mathcal{Q}$ is the interview question space. In the initial session, we send $u_1$, $\boldsymbol{\theta}_0$, and $\rho_0$ to $\mathcal{LM}^q$ to obtain $q_1 \in \mathcal{Q}$. That is

$$q_1 = \mathcal{LM}^q(u_1; \boldsymbol{\theta}_0, \rho_0) \tag{1}$$

Note that the output of $\mathcal{LM}^q$ is stochastic if the temperature parameter in $\theta_0$ is greater than 0. However, the output is still a specific question $q_1$ rather than a probability distribution. In this sense, we can also describe Eqn. (1) as $q_1 \sim \Pr(q_1 | u_1, \boldsymbol{\theta}_0, \rho_0)$. Additionally, this step degenerates into simply $q_1 = u_1$ if we define a specific fixed question that will be asked without any adjustment or random perturbation by prompting the agent to, say, use a variation of the question specified in $u_1$.

Let the participant's response (through a UI) to $q_1 \in \mathcal{Q}$ be $r_1 \in \mathcal{R}$. We then send $r_1$ along with $q_1$ and $u_1'$ to an LLM with an adjusted system setting so that $\mathcal{LM}^s: \mathcal{R}, \mathcal{Q}, \mathcal{U}, \mathrm{P} \to \{0,1\}$. $u_1'$ is a follow-up prompt (template) regarding $u_1$. For instance, $u_1'$ can be the following prompt template: "Decide if [$\mathcal{R}$] responds to [$\mathcal{Q}$]." That is, $\mathcal{LM}^s$ determines whether $r_1$ responds to $q_1$ sufficiently. We define the output as $\xi_1 \in \{0,1\}$. That is

$$\xi_1 = \mathcal{LM}^s(r_1, q_1, u_1' | \boldsymbol{\theta}_0, \rho_0) \tag{2}$$

Next, we define a conditional logic $f$, which triggers the next round (about the production of $q_2$) if $\xi_1$ is 1; otherwise, it triggers the clarification request $q_1^c$ through Eq. (3):

$$q_1^c = \mathcal{LM}^q(u_1^c[r_1, q_1]) \tag{3}$$

where $u_1^c[r_1, q_1]$ indicates that the prompt template $u_1^c$ has become a completed (ready-to-use) prompt with filled information $r_1$ and $q_1$. The new response to $q_1^c$ is stored as $r_1^c$. The $r_1$ and $r_1^c$ together form the new response $r_1$. We then send the relevant historical conversation ($r_1$, $q_1$, $u_1'$ in this case) to $\mathcal{LM}^s$ to determine once more whether the cumulative response thus far about $q_1$ is sufficient (i.e., whether $\xi_1 = 1$). To simplify the notation, we define a tuple $y = (r_1, q_1)$. Note that, the new $r_1$ is formed by appending new clarification responses to $y$. Note that $y$ is a tuple with variable length, as the response to Question 1 may or may not induce one or more follow-up questions. This update process continues until the clarification is considered sufficient by $\mathcal{LM}^s$ or the process reaches a pre-determined maximum allowable clarification number (e.g., only ask a participant up to twice for clarification).

We use $\mathcal{LM}^s$ (again but with adjusted system setting) to extract information from $y$ about the key information that needs to be obtained from this round of conversation; this extraction not only reduces token consumption for memorizing prior rounds of conversation with the present participant but also speeds up the agent's response time in the following rounds. Suppose we denote the extracted information about the response to $q_1$ as $x_k \in \chi$, $k = \{1,2, \dots, K\}$, which is obtained as shown in Eq. (4):

$$x_k = \mathcal{LM}^s(y_1, q_1, u_1^k | \boldsymbol{\theta}_0, \rho_0) \tag{4}$$

The $K$ pieces of information, as placeholders, are predetermined in the S&I design phase based on the overall objectives of the survey or interview. For example, $x_1$ might store the participant's age, $x_2$ might store the participant's gender, and $x_3$ might store whether the participant uses a certain product or a mode of transportation more than three times a week. All $x_k$ ($\in \mathcal{X}$) form an $K$–dimensional vector $\boldsymbol{x}$. $u_1^k$ is the prompt (or prompt template) for instructing the agent to extract information from $y_1, q_1$ to $x_k$. This process continues until $\xi_1 = 1$ or the maximum number of allowed iterations is reached. $\xi_1 = 1$ also means that all the necessary information about Question 1 has been obtained. Fig. 3 illustrates the proposed process for the initial session, where the agent only asks a follow-up question once (and moves on to the next question regardless of whether the participant sufficiently clarifies his/her response). Note that, if the question is about a participant's personal information, additional care should be taken (e.g., forbid the use of commercial LLM for implementing $\mathcal{LM}^s$) to remove privacy concerns.

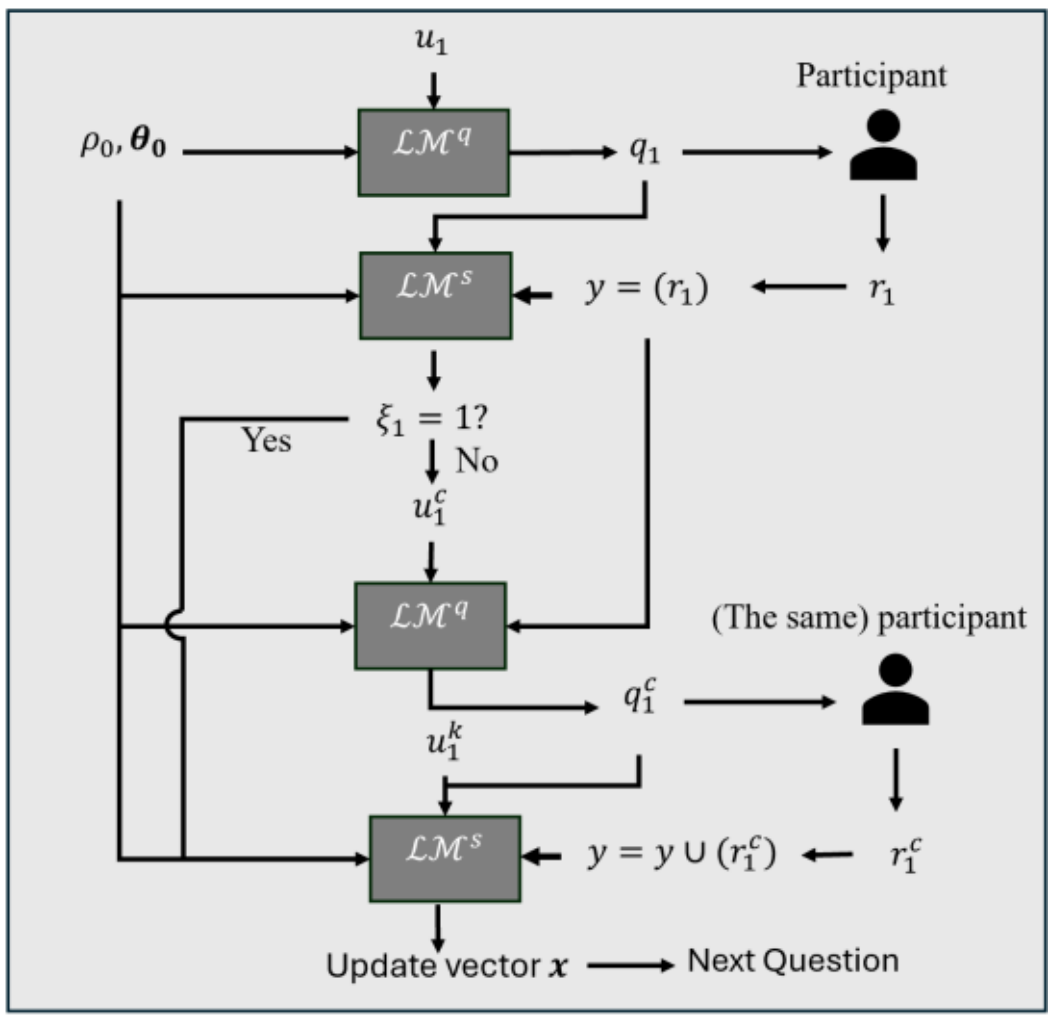

**Fig. 3** Illustration of an information flow for the first question in a three-question semi-structured interview, where the AI agent asks for clarification only once.

Next, the agent enters the second session (about Question 2). This session has a similar initialization process. However, this time, we might need to incorporate information from the prior round(s) through $x_k$ if the specifics of Question 2 depend on certain information related to the response to Question 1. That is

$$q_2 = \mathcal{LM}^q(u_2[x_k]; \boldsymbol{\theta}_0, \rho_0) \tag{5}$$

where $u_2[x_k]$ denotes the prompt that combines prompt template $u_2$ with the information $x_k$ form the prior section (question). We might further define $\boldsymbol{\theta}_2$ and $\rho_2$ as the parameters and template, i.e., $u_2 = u_2[x_k; \boldsymbol{\theta}_2, \rho_2]$, to further increase the flexibility of $u_2$. The same method was used for the other interview questions. The participant's response to $q_2$ is denoted as $r_2$, and will be sent to $\mathcal{LM}^s$ again, along with $q_2, u_2, x_k$, with output $\xi_2$. Again, this process continues until $\xi_2 = 1$ (or the maximum number of allowed iterations is reached). Note that utilizing the entire $y_1$ (instead of $x_k$) might produce richer information for $\mathcal{LM}^q$, but it might consume significantly more LLM tokens and longer LLM response delays.

Once we enter the third session (for Question 3), we have a similar initialization process. However, this time, we might need to incorporate information from the prior two rounds through $x_k$ and $x_{k'}$.

$$q_3 = \mathcal{LM}^q(u_3[x_k, x_{k'}]; \boldsymbol{\theta_0}, \rho_0) \tag{6}$$

We can continue this process with more than just three questions. Furthermore, using the logic function $f$, we can also branch out different questions on the basis of the specific user background and state. Additionally, during the conversation, $\boldsymbol{\theta}_0$ and $\rho_0$ can be adjustable, but the boundary conditions and the resultant impact on reliability deserve further research.

We can generalize the above formulation for unstructured interviews. Instead of storing specific $u$ for each pre-termined question or question instruction, we can store the overall objective of the interview in $u$, and instruct the agent to obtain the objective by asking individual questions. Once the participant responds, the agent summarizes the response and may ask for clarification. If there is no need to clarify, the agent will compose the next question(s) with the prompt containing the overall goal of the interview until the goal is deemed reached by the agent. The agent might be even further augmented by external knowledge databases or knowledge bases that are gradually formed through S&Is, enabling the agent to continuously learn and improve.

### 3.3. Sensitivity analysis and fine-tuning of prompts and parameters

With the parameterized description of a LLM-powered S&I process, we can further describe the methods of conducting sensitivity analysis of system settings ($\boldsymbol{\theta}_0$ and $\rho_0$) and question-specific $u_i$. The sensitivity of the LLM outputs to inputs and parameters is analyzed by varying the contents and wordings of the system prompt, parameters and question-specific prompt templates. This analysis is crucial for understanding how different linguistic formulations affect the performance of the agent and how to improve it. Let $\boldsymbol{\theta} = \{\theta_0, \theta_1, \dots, \theta_i, \dots, \theta_I\}, \boldsymbol{\theta} \in \Theta$, be a set of $I + 1$ parameter vectors (each element might also be a vector). The marginal sensitivity of $q_i$ to $\rho_i$ and $\theta_i$ can be presented as $\Delta q_i / \Delta \theta_i$ and $\Delta q_i / \Delta \rho_i$, respectively, with $\mathbb{E}(\Delta q_i / \Delta \theta_i)$ and $\mathbb{E}(\Delta q_i / \Delta \rho_i)$ as the average effects. Hence, the average effective of marginal variation of $r_i$ with respect to $\theta_i$ can be expressed as $\mathbb{E}(\Delta r_i / \Delta q_i \cdot \Delta q_i / \Delta \theta_i)$, or simply $\delta^i_{\theta:r}$. Similar principles applies to the variation of $u_i$. Since not every variation of $q_i$ is a suitable or appropriate question, we define a dummy variable $\sigma_i$ to indicate whether adjusted $q_i$ is (still) a valid question. Therefore, the expectation of the impact of a marginal variation of $u_i$ can be described as can be represented as

$$\delta_{u:r}^{i} = \mathbb{E}\left(\sigma_i \cdot \frac{\Delta r_i}{\Delta q_i} \cdot \frac{\Delta q_i}{\Delta u_i}\right) \tag{7}$$

where $i \geq 1$. A high sensitivity indicates that the function is highly responsive to changes in the corresponding input or parameter, which may require careful control or robust design. On the other hand, low sensitivity suggests that the participant's response is relatively insensitive to the variation of the system setting (and the associated question), allowing for more flexibility in design and operation. Similarly, we can obtain the expected sensitivity of one system parameter or variable on the output. For example, the impact of a marginal variation of $\rho_i$ on $x$ can be represented as $\delta_{\rho:x}^{i}$, which follows a similar principle. Again, as not all the directions of perturbation are feasible (e.g., changing one letter or word in a prompt sentence might lead to an incomprehensible or grammatically incorrect sentence), the equation above might not be applicable, not to mention that the differentials might not exist. Therefore, $\sigma_i$ is a dummy variable—when the direction along $\theta_i$ is differentiable for $u_i$, we have $\sigma_i = 1$; otherwise, it is zero. We can continue this evaluation process to obtain the impact of varying $u_i$ on $y_j$ (and $x_k$ that corresponds to the $j$th question), where $j > i$.

Owing to the complexity of the internal mechanism of an LLM, even if we have complete information about the model architecture and parameters, it might still be impractical to obtain a complete set of sensitivity analysis. However, the formulations above still provide valuable insights into how the sensitivity of prompting information on the output can be evaluated. The resulting insights might further help enhance the design of conversational flows to improve elicitation quality.

## 4. Field Experiments

We present three concrete cases in which the proposed architecture and process flow for conducting S&Is are utilized. The user interfaces (dialog boxes) for all three experiments are accessible through smartphones and computer web browsers. The proposed modular approach is platform-agnostic. As we are writing the paper, various online platforms are available for AI agent development and deployment. To clarify the role of the proposed modular approach and platforms for AI agent development and deployment, we can draw an analogy to designing a bridge with computer-aided design (CAD) software/platform. Just as CAD software supports and facilitates an engineer in creating a complex structure (the bridge), a platform for AI agent development and deployment can help realize and implement our proposed modular approach in building an AI agent for S&Is. Criticizing the methodological contribution of our approach because it utilizes AI development platforms would be akin to attributing all credit for a bridge's design to the CAD software. The proposed approach is also module-agnostic. That is, any module shown in Fig. 1 is replaceable as long as the new module can serve the same function. For example, one can choose any open-source LLM for the on-premise LLM module. For another example, one can design their own user interface (UI module) using a preferred programming language.

The online LLMs used in the first, second, and third experiments are from GPT-4, GPT-4o, and Claude 3-SonNet, respectively. In addition to using Llama 3.2 for the on-premise LLM component for privacy protection, we implemented a set of additional measures to ensure the protection of participants' data. First, we make it optional for the participants to decide whether they would like to share their (potentially) identifiable information. Second, even if they share their (potentially) identifiable information such as name, workplace, and email address, we employed anonymized techniques to dissociate any personally identifiable information in data processing and in released results. For home location information, we only ask participants for the first three (of the six) digits of the postal code of their home address in Canada. The response goes through an on-premise LLM to detect any privacy concerns. Only then will the process manager module consider whether to send the information to cloud-based LLMs for further processing. The on-premise LLM might also suggest that the process manager hide or modify certain sensitive information before being sent to an online commercial LLM. Furthermore, we prioritized secure data storage practices. The collected data were stored on dedicated hardware that featured security protections provided by McGill University. Only authorized researchers and supervisors were granted access to this hardware, and stringent access control mechanisms were in place to safeguard against unauthorized access or breaches. IP addresses are not associated with usernames or user IDs, and they are only used to identify when a new request is needed during the operation of an AI agent, whether to continue from the previous chat (if any) or launch a completely new conversation.

Approval for all ethical and experimental procedures and protocols was granted by the Research Ethics Board (REB) office at McGill University and by the Committee on the Use of Human Experimental Subjects (COUHES) at the Massachusetts Institute of Technology.

### 4.1. Experiment 1: Resident travel preference elicitation

In the first study, we developed an agent and employed it to elicit residents' preferences regarding their travel modes in response to various factors, especially weather conditions. The aim of this study is to understand how different weather scenarios influence residents' choice of travel mode so that policymakers can determine better incentives for promoting sustainable means of transportation. Some of the questions are aided by weather images generated by the agent or pulled by the agent from an image database. During the survey, the participants were asked a series of questions (with discrete options generated by the agent) about their travel preferences under different weather conditions. The agent gradually identifies participants by categories, primarily on the basis of their dominant travel mode(s) and asks increasingly customized questions oriented toward the goal of understanding a participant’s mode preference under different weather conditions. For example, by knowing whether a participant is an avid bike user, the following questions will focus on the conditions under which they would consider using alternative modes (other than bikes). In addition to the text display, users also have the option to use voice input to make their choices. The voice is then further processed to infer the intent behind the participant’s choices. If none of the options are matched, the agent will ask for clarification. In some questions, the agent also presents images of various weather conditions to provide participants with a concrete sense of the type of weather being discussed.

On average, 19 out of the 21 questions (not all the questions are asked for a specific participant) in this experiment are discrete choices, with the rest of the questions being open-ended. Some discrete choice questions have the option “Other”; if chosen, the agent will ask the participant to type in the specifics and determine whether the response is relevant to the question (and ask for clarification once). In open-ended questions, the agent summarizes the responses and might ask for clarification (only once) if the agent deems the response irrelevant or incomplete. After the response to a clarification question, the participant will have a chance to decide if they would like to further clarify or go to the next question.

The participants interacted with the agent through a simple dialog interface, as shown in Fig. 4a, where they could view the current conversation and access the full history of exchanges with the agent. This interface was designed for ease of use and is accessible via any standard web browser, whether on a laptop or mobile device. To enhance clarity in certain scenarios, such as describing weather conditions, the agent primarily generates images rather than text. For example, when participants are asked about their preferred means (modes) of transportation, the agent employs generated images to illustrate different weather conditions, as shown in Figs. 4b and 4c. To ensure reliability, we allow LLMs to pre-generate a set of images and store in an image database, but one can also let LLMs generate images (with a certain level of randomness) on the fly to create more diverse weather scenarios on the basis of the responses provided in earlier phases of the conversation.

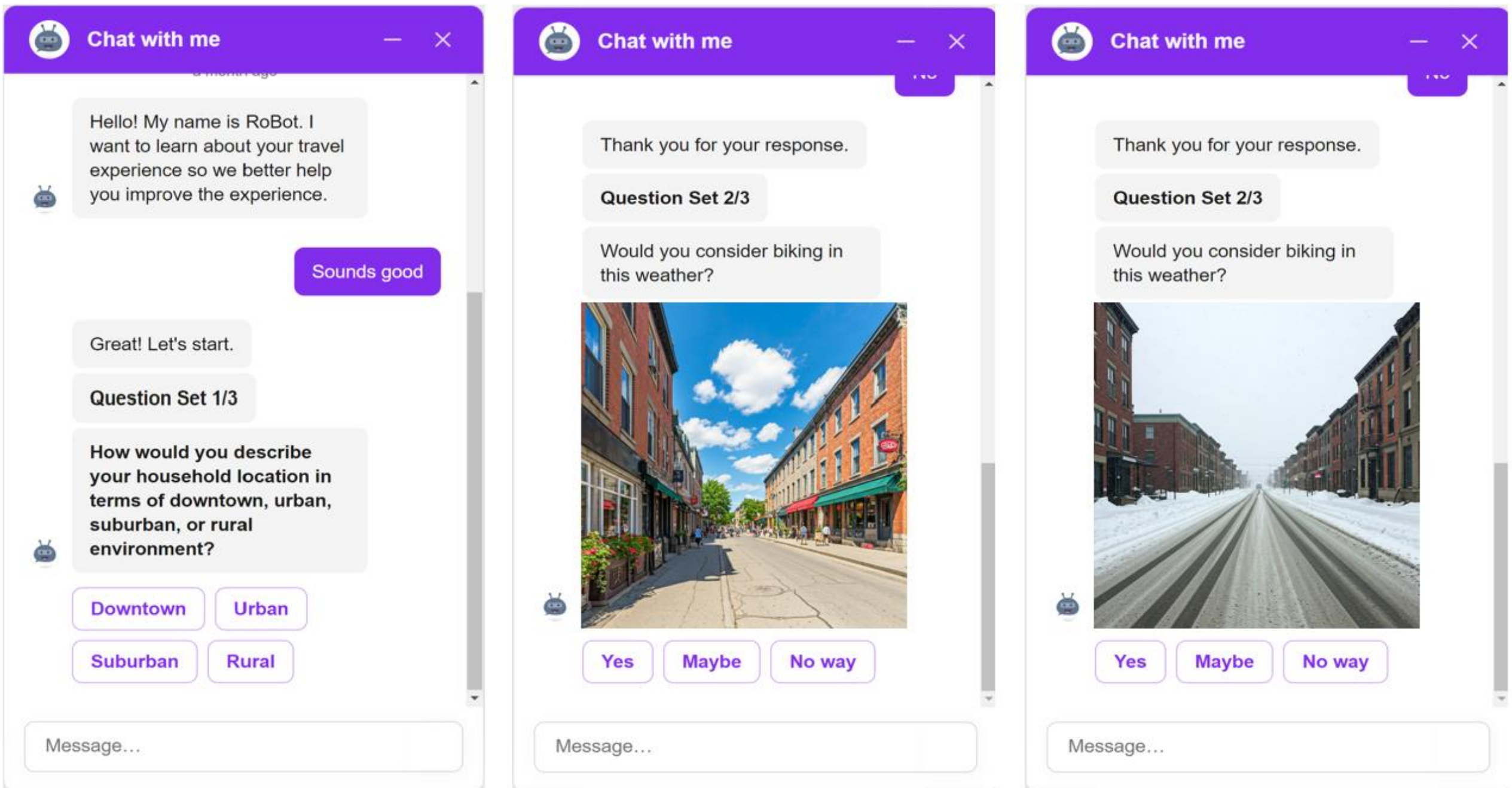

**(a)** **(b)** **(c)**

**Fig. 4** Interface with the AI agent during a survey on residents' travel preferences. Both options and images are AI generated with customization. English and French versions are available.

Conditional logic was embedded in the dialog flow to address changes in participants' transportation choices due to shifting weather conditions. For example, if a participant indicated that they usually bike during a sunny summer day, the agent could ask a follow-up question, such as "Would you still bike if there is moderate rain, as shown in the image below?", with a picture depicting weather with moderate rain. This logic was especially useful for addressing less intuitive decisions, such as switching from driving to biking when there is rain, to not only dig deep into the rationale of mode choice but also ensure that participants fully understood the question. To explore the impact of varying conversational styles, the agent randomly selects different voices and questions phrasing from a pool of pre-defined question variations. This ensured that each survey session featured a diverse set of conversational approaches.

We distributed the agent through a Quick Response (QR) code along with a website address (URL) to 210 undergraduate and 47 graduate students at McGill University, most of whom were from the Department of Civil Engineering. No participation incentives were provided, so the participants who completed the survey might be biased towards those that are interested in transportation studies or related topics. In total, we received 184 responses, 103 of which completed the survey. The collected data were then examined to assess the impact of weather conditions on mode choice (or mode shift). The key factors considered in the analysis included weather and seasonal conditions (rain, snow, sunshine, and wind), travel time (short, medium, and long), transport mode (walking, biking, public transit, and driving), and safety concerns (traffic safety and personal security). After the interview, we also convened the 15 graduate students to ensure that the responses were truthful to their selections, and all of them confirmed that the response documented by the agent was correct. BIXI is Montreal's public bike-sharing system that provides users with access to regular pedal-powered and electric-assisted bicycles through a network of docking stations across the city. Fig. 5 illustrates the (unweighted) counts of primary travel model by season (summer and winter). While detailed data analysis and policy recommendations are beyond the scope of this paper, the preliminary findings reveal a clear and consistent pattern. Future work will involve collecting data from a broader and more diverse participant pool, conducting an in-depth analysis (such as developing discrete choice models to interpret the factors influencing travel mode choices), and performing statistical tests to assess the significance of each factor and subfactor. These results and insights will be presented in future publication.

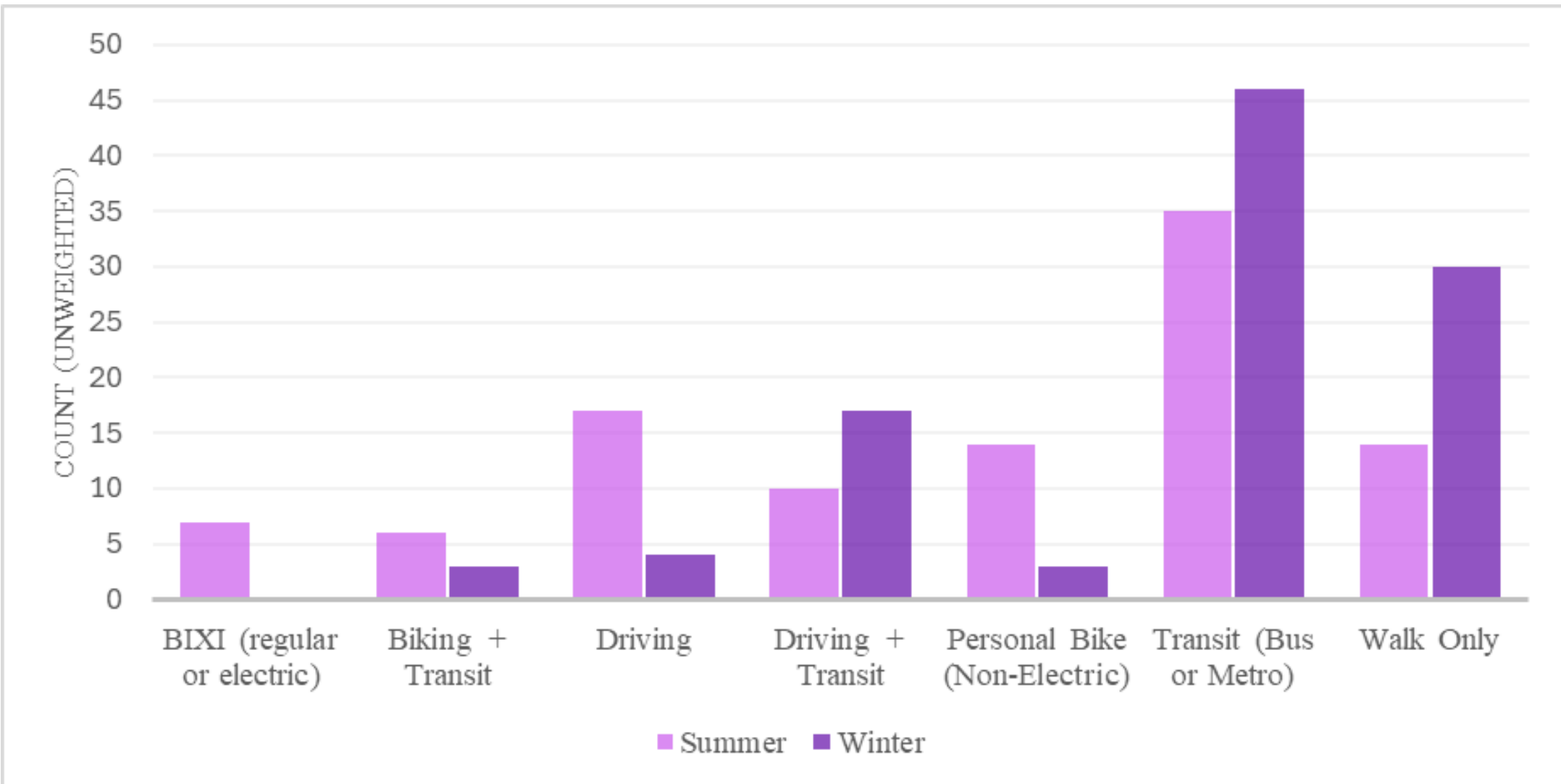

**Fig. 5** Primary mode of travel in summers and winters.

This demonstration confirms that the agent is a valuable tool for capturing residents' travel preferences under various conditions with low cost and high efficiency. Additionally, it is worth noticing that before we deployed the AI-aided survey agent, we distributed a similar survey using a regular web-based questionnaire (using Qualtrics) to about 20 students (also no incentives were provided), and only three students completed the survey. Therefore, the relatively high response and completion rates suggest that (conversational) AI agents might promote participation than conventional S&I methods.

### 4.2. Experiment 2: Public perception for infrastructure project

The second study develops and deploys an AI agent using the proposed modular approach for understanding the public perception of a recently (September 2024) implemented pedestrian crosswalk lighting system in the City of Candiac, Quebec, Canada, which is considered the first of its kind in North America (CBC News, 2024). The lighting system is shown in Fig. 6. The targeted participants of the students at the St-Mark's Elementary School adjacent to the new lighting system.

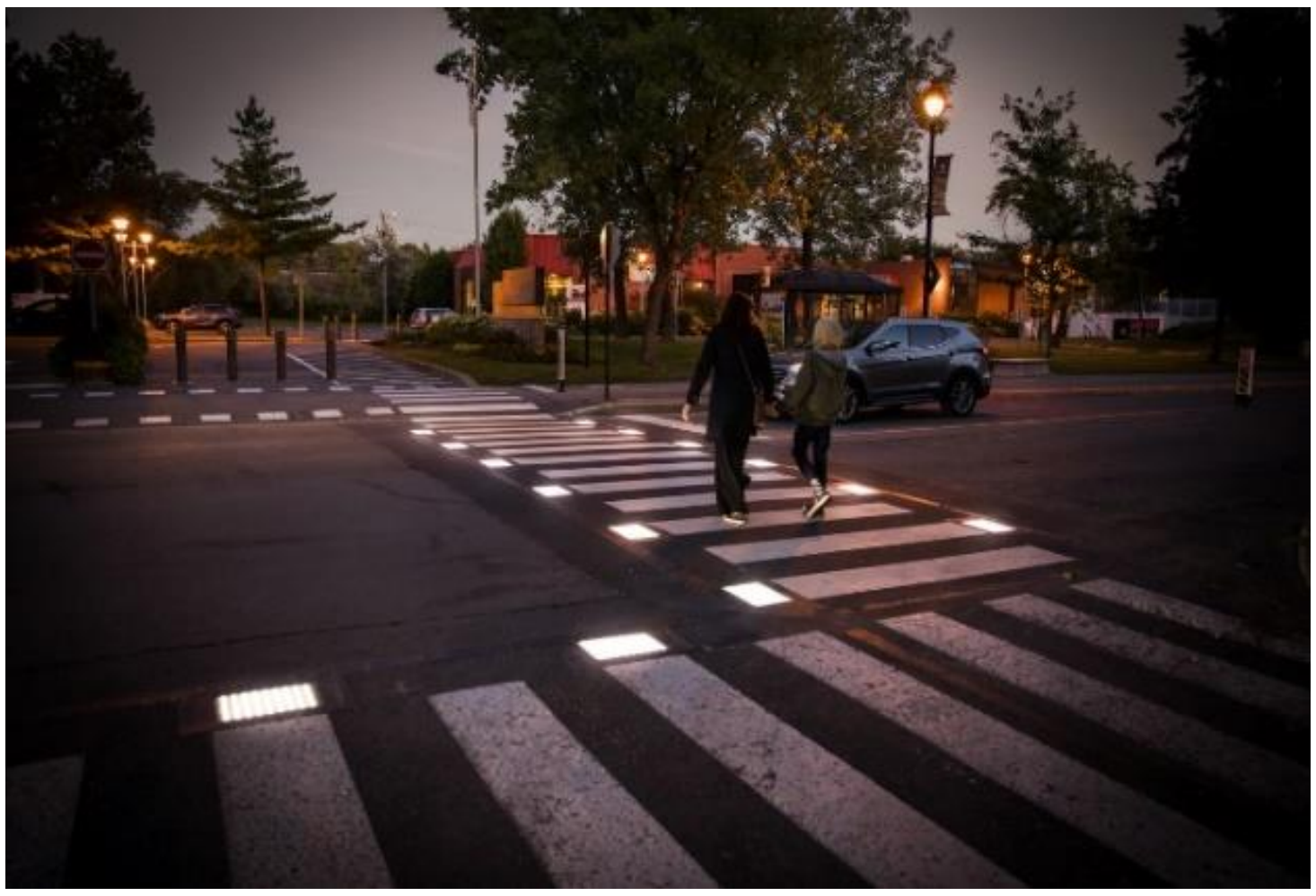

**Fig. 6** Newly installed lighting system in Candia, Quebec, Canada, for improving intersection traffic safety. The image is adopted with permission from the Ville de Candiac [Source].

The City of Candiac tested the AI agent and distributed it among residents (a major portion is the parents from the nearby elementary school), with an email containing a URL and a QR code to the agent. Participants can directly click the link on their smartphone or use their laptop or tableau to scan the QR code. The interface of the agent is shown in Fig. 7. For comparison, we also developed a web-based questionnaire version of the same set of questions. The difference is that the web-based version does not have AI assistance and the ability to be conversational, such as paraphrasing participants' responses to ensure clear understanding and asking for clarifications. The web-based form also does not allow voice-based responses. Our system randomly distributed 80% of the participants for the web-based questionnaire and the other 20% for the AI agent. In both cases, the survey starts by informing participants about the study's objective, the agent's functionalities, and the data privacy measures. The participants can choose to use French or English modes. Note that, in the case of AI agent, there are no two separate process flows for each language; instead, it is one single process flow, where the LLM can take French or English as inputs and produce output using the language specified in the associated prompt parameterized by a variable (about what language to use) based on the participant's response to the initial question about the preferred language. The agent also allows the participants to interact through voice commands, which was particularly beneficial for those who preferred speaking over typing. In discrete choice questions, the agent attempts to match the voice answer with the options available. If there is no match, the agent will ask for clarification. In open-ended questions, the agent attempts to summarize or paraphrase the participants' responses and asks for clarifications when necessary to ensure data accuracy. If the agent considers a response to be challenging to summarize or paraphrase (likely due to response irrelevance), the agent asks the participant to clarify with more detail and paraphrase the participant's response to ensure its understanding.

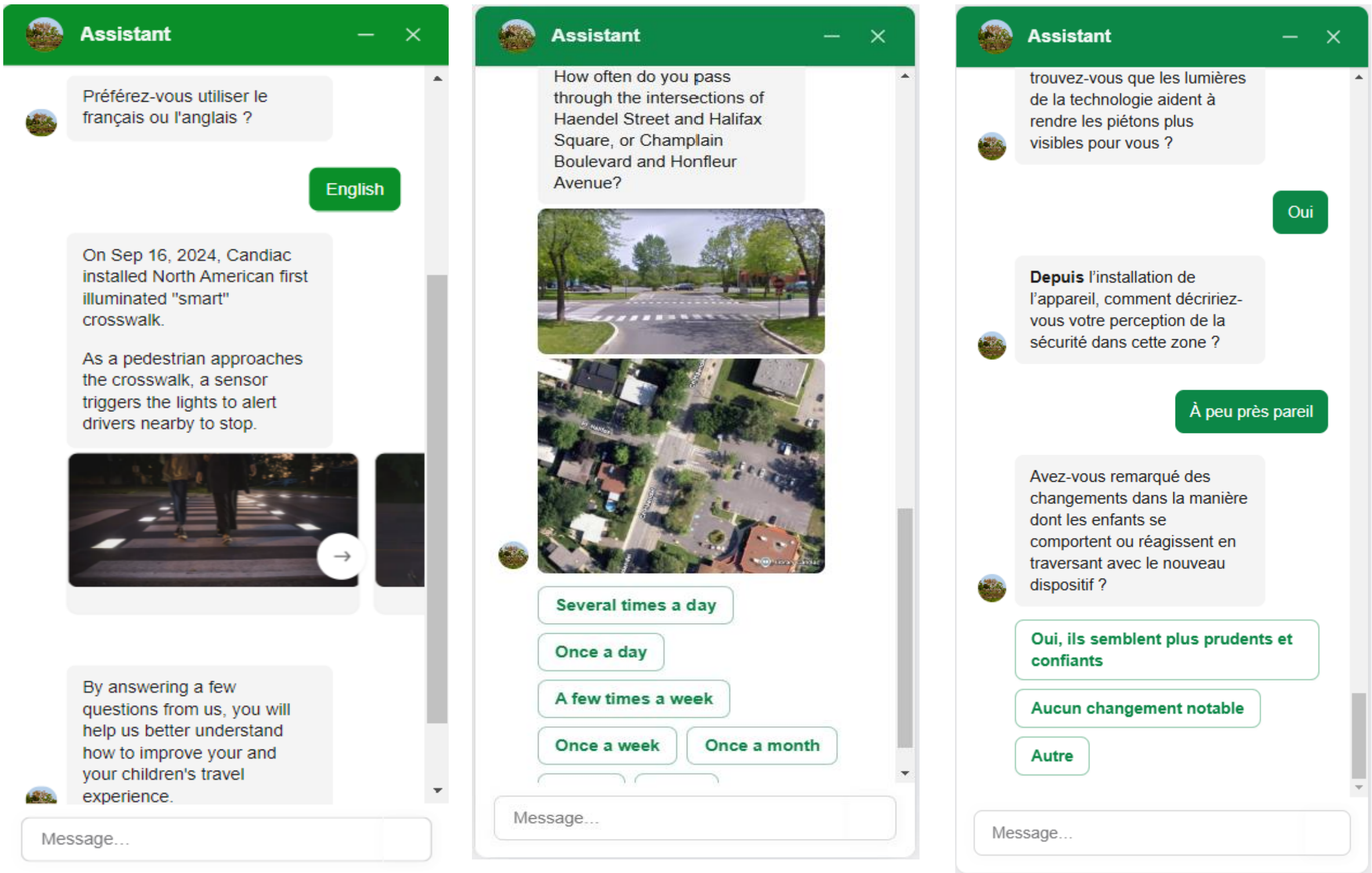

**(a)** **(b)** **(c)**

**Fig. 7** Agent interface during an interview. The agent can use English or French to communicate with the participants and follow up for further clarifications.

The process consists of three main phases (each phase has approximately 7 questions): (1) collect information about the travel mode, frequency, and time of day for the participants to pass through the intersections or send their child (children) to the school and whether they have noticed the lighting technology; (2) based on the participants' responses, the agent will either ask whether the parents think the technology is effective in terms of safety improvement or explain the technology is about; and (3) elicit opinions (including both discrete choice and open-ended questions) about any other thoughts and suggestions and whether the parents think the technology should be installed in more intersections in the region.

A total of 117 participants started the survey with the agent, and 29 entered the web-based survey (without AI enhancement). We did not provide participation incentives in either formats except that we remind the participants that their responses will be helpful for the city to better understand how to improve their travel experience in the future. Of the participants who started the survey with the agent, 78 (approximately 66.7%) completed the survey with sufficient quality; Of the participants who entered the web-based survey, 11 (approximately 37.9%) completed the web-based survey with sufficient quality. The difference in survey completion percentages suggests that the conversational nature of the AI agent might promote more participation than the web-based form (not conclusive due to the relatively small sample size).

We further prompted the AI agent to only elicit the opinions but also postprocess the responses into a standard data analytics (spreadsheet) format. We manually checked the organized data and found a high level of accuracy. Since this section primarily aims to demonstrate the efficacy of the agent using the modular approach, and we have not yet obtained permission to disclose the full survey results, we present only preliminary findings in Fig. 8 for demonstration purposes. Fig. 8a illustrates that 85% of participants passed through the intersection at least once a month in the past year. Fig. 8b shows that approximately 74% have used the smart crosswalk, while Fig. 8c indicates that over 76% traversed the intersection on foot or by bike, though some may have also done so by car. Fig. 8d highlights that more than 62% frequently cross the intersection at night. Regarding perceptions of the new lighting system's impact on safety, Fig. 8e shows that 50% believe it enhances safety, while 40% do not. Among those who walked or biked through the intersection with the newly installed lighting system, Fig. 8f reveals that 39% found drivers' behavior toward them to be more considerate. Fig. 8g indicates that 88% of respondents believe the new lighting system improves nighttime visibility in regular weather conditions, and Fig. 8h shows that over half of the participants find that it enhances visibility in snowy conditions. Finally, Fig. 8i suggests that about 55% of respondents think similar lighting systems could improve safety in other locations. Overall, these preliminary findings suggest that most participants familiar with the newly installed crosswalk lighting system hold neutral to positive views about the project. However, as this lighting technology is still relatively new, it remains uncertain whether perceptions will shift with increased familiarity over time. Future studies are needed to assess the long-term stability of these initial trends, and we leave the comparison between responses obtained from the agent and those from the web-based questionnaire for future research.

The experiment demonstrated the effectiveness of the LLM-enabled chatbot in conducting public outreach for infrastructure projects. Approximately 20% of the participants chose to use voice, and 83% chose French, which suggests that the agent's voice capabilities and bilingual functions are valuable for encouraging S&I participants who otherwise would not participate. We also noticed that in approximately 4% of the discrete-choice questions and 30% of the open-ended questions, the agent asked for clarifications (we set the maximum number of clarification requests to two), and we did find that the clarification provided clearer and more definite answers.

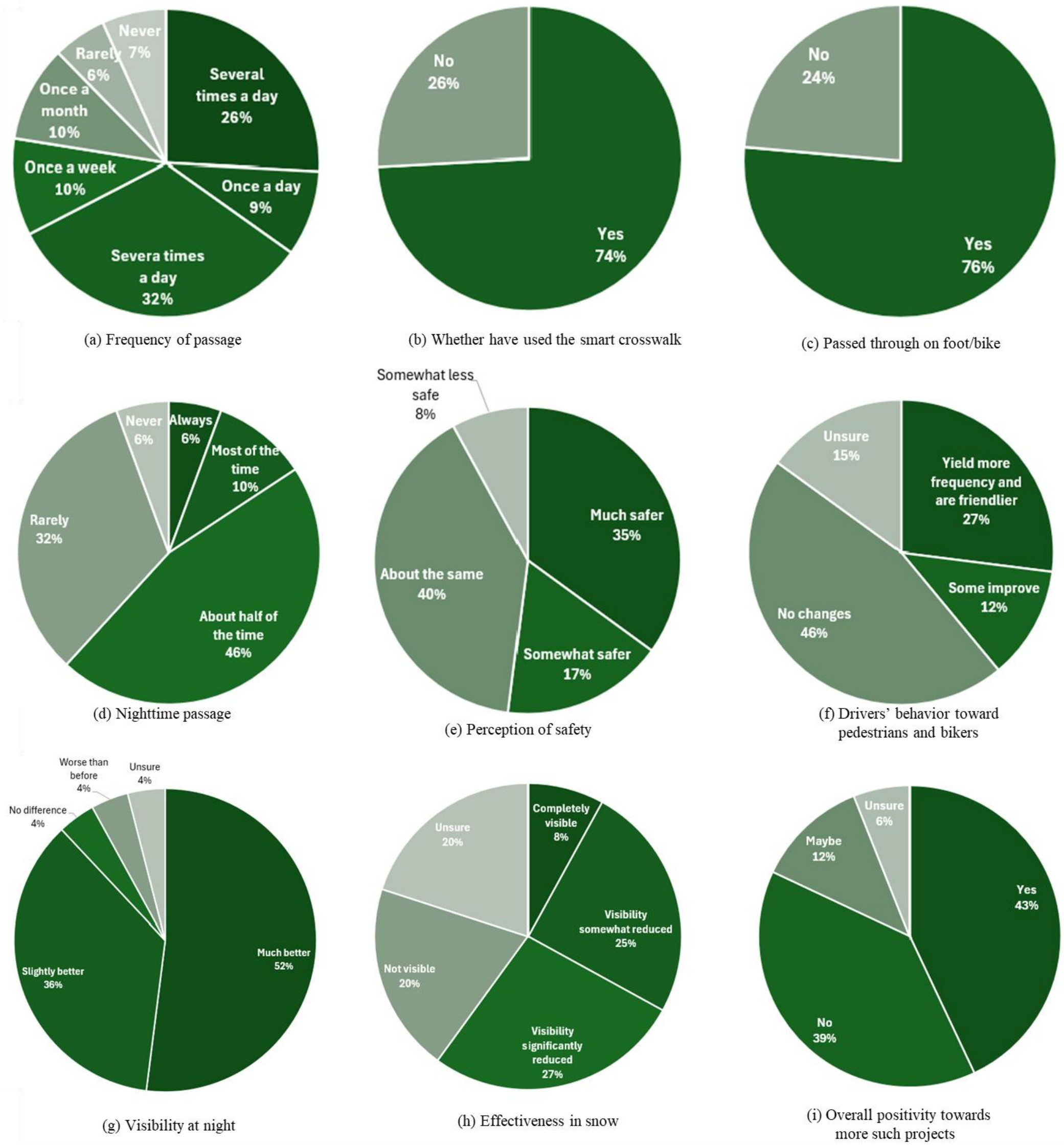

**Fig. 8** Preliminary survey results (unweighted) on usages and perceptions of the smart crosswalk lighting system

### 4.3. Experiment 3: Expert opinion elicitation (open-ended questions)

In the third study, we designed and deployed an AI agent via the proposed modular approach to conduct a semi-structured interview with transportation professionals, asking for their opinions on the current and future potentials of LLMs in the field of transportation and urban mobility. The interview consists of five open-ended questions, each of which might be associated with up to two follow-up questions for clarification. The agent was distributed through different channels, including the Massachusetts Institute of Technology (MIT) Mobility Initiative email list, LinkedIn announcements, and some additional tests with the students and researcher associates at MIT and McGill University. Overall, we received approximately 53 high-quality responses from the two versions of the agent, which have nearly the same set of five questions and only minor wording adjustments and temperature parameter (for LLM output randomness) differences. We leave the study of the impact of these minor variations on responses to future research. As 41 of these participants left their names and contact information, we know that they are mainly from universities, consulting companies, transit operation agencies, and regional planning agencies in North America, Europe, and East

Asia. Although the agent uses English only, it does not show problems in responding to participants who answered the questions in German (one participant) or Korean (one participant). All other users used English in their responses.

The main objective of this experiment was to evaluate the agent's ability to elicit expert opinions effectively and efficiently. The agent followed a pre-determined sequence of topic questions. After responding to each question, the agent paraphrases and summarizes the participant's response in conjunction with prior questions and responses to ensure that the agent understood and interpreted the response correctly in the context of responses that the participants already gave to prior questions. If major missing elements in a question were identified in a participant’s response, the agent requests clarifications. Additionally, the chatbot provided the participant with an opportunity to add clarifications voluntarily. This procedure aims to increase the accuracy and reliability of the collected data. To ensure that the agent understands common jargon and terminology in the field of transportation engineering and planning, we incorporated a knowledge base from the Transportation Research Thesaurus (https://trt.trb.org/), retrieved via the RAG technique. We found that this approach resulted in a marginal improvement in understanding (when a participant uses technical terms), albeit with some minor slowdown in the agent’s response time.

Fig. 9a shows the interface and the initial state when a chatbot is launched (either on a web browser or a smartphone). Figs. 9b and 9c show the agent responses during conversations, where the agent asks for clarification, as it considers the response from the participant incomplete.

As exemplified in the Appendix, the agent demonstrates a high level of accuracy in paraphrasing responses and requesting clarifications. Multiple participants reported near the end of the interview that the paraphrasing feature helped them reflect on their responses and provide more precise answers. Compared with the five in-person interviews before the agent was developed, the efficiency of the agent-mediated interviews was greater, with slightly shorter average interview times (although only a text-based conversation was allowed by this agent), consistent data quality, and significantly more condensed meaningful information (perhaps because the participants thought more carefully through typing their opinions).

Before deployment, we tested a wide range of potential erratic behaviors, such as answering a question with an irrelevant response and simply asking the same question back. During deployment, one participant indeed attempted to leave the agenda by making an irrelevant request at the beginning of the interview – asking the agent to tell a joke. These erratic behaviors are responded to with an apologetic statement of not being able to answer and then reiterate to the original interview question. In terms of the reliability of the LLM responses, although we did not observe any hallucination phenomena (i.e., an LLM making up facts), we acknowledge such risks. On the one hand, we further prompt the LLMs to not fabricate factors but simply admit that they are not known. On the other hand, we suggest that future research, including better utilization of RAG technologies, address this issue to ensure complete reliability.

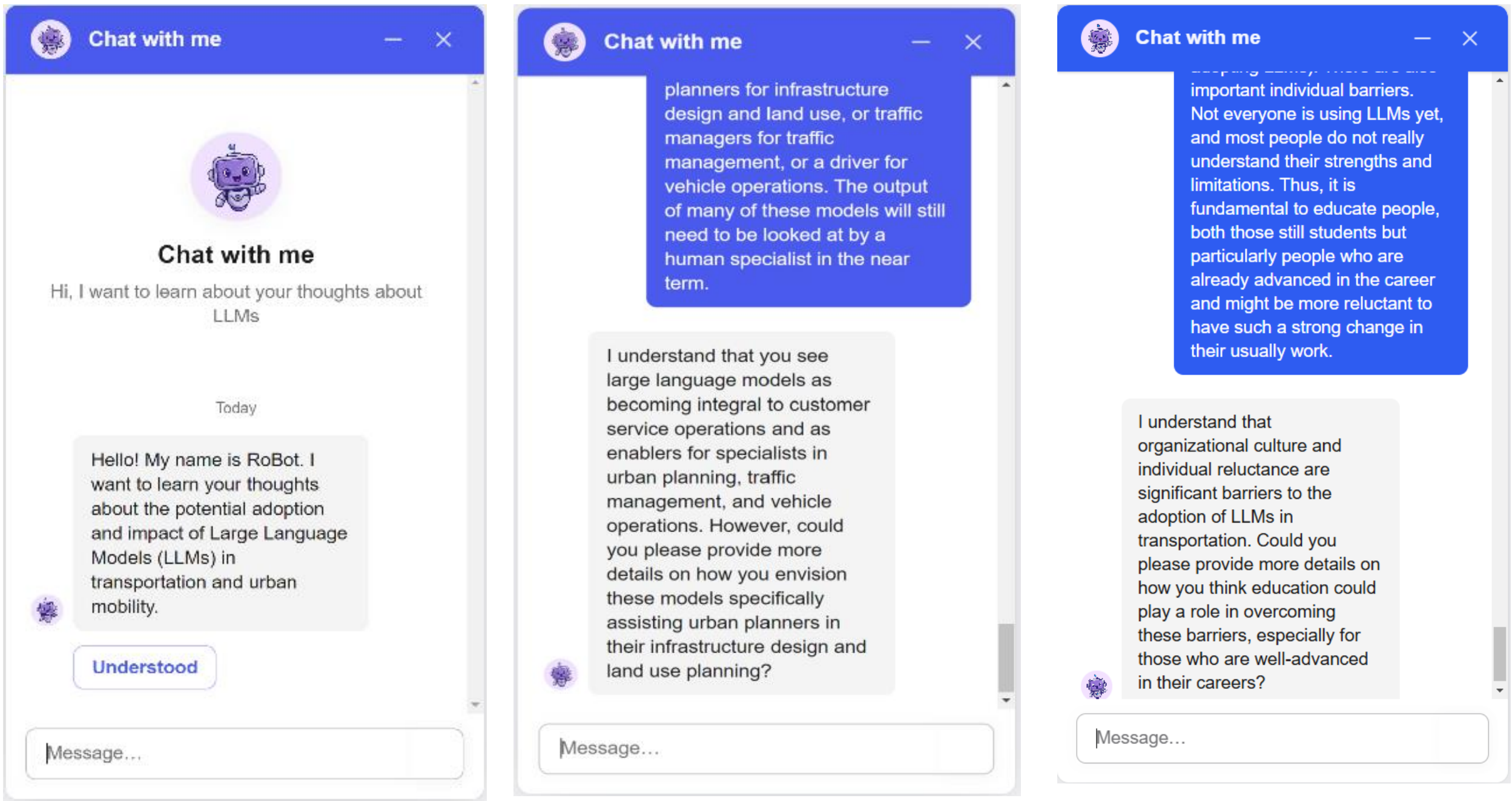

**(a)** **(b)** **(c)**

**Fig. 9** User interface for conducting semi-structured interviews on experts' opinions on LLM for the future transportation industry.

**Table 2** Key phases in participants' responses in experiment 3.

| Question | Common keywords from the response |
|---|---|
| Q1. Specific applications of LLMs in workplace. | R1. Drafting emails, refine reports, asking to clarify technical jargons. |
| Q2a. Benefits of LLMs. | R2a. More efficient workflow, low cost, lower work burden. |
| Q2b. Challenges of LLMs. | R2b. Information inaccuracy, unclear sources/citations, worry that they might be replaced one day. |
| Q3a. Future LLM applications and impacts (next 1–3 years). | R3a. More commonly used among colleagues, automatic email responses with daily summaries and approval requests. Automatic reports/documents generations with periodic update. Integration with traveler information systems. |
| Q3b. Future LLM applications and impacts (next 5–10 years). | R3b. Integration with autonomous vehicles, replacing human engineers, planners and researchers, better and more adaptable traffic control, recommending transportation plans but are challenging to be humanly understandable. |
| Q4. Main barriers for beneficial adoption of LLMs. | R4. Technology capabilities, biased regulations, radical ideations, conservative people, risks of cyberattacks. |
| Q5. Visions (if no limit) | R5. Real-time communication, personalized route-suggestions, public transit optimization, cost-benefit analysis, accountability, equity improvement/worsening, last-mile delivery update, data privacy concerns, seamless system integration. |

We performed a keyword analysis to identify common themes and issues mentioned by the participants. The results are summarized in Table 2, which lists the final set of identified issues, objectives, performance metrics, and decision weights. The original transcripts already contain a summary and paraphrase of each participant's response to each question. Therefore, we employed an LLM to further streamline and summarize the responses by question. During and after this procedure, we manually checked the samples to ensure that the summaries did not show signs of hallucination and accurately reflected the participants' opinions. The results suggest that the agent's ability to ask for clarification and follow-up questions significantly improves the quality of the response collected.

## 5. Discussion and future work

The modular approach and the parameterized conversational process it enables offer notable advantages, particularly in standardizing large-scale data collection, providing multi-modal real-time interactions with domain expertise, and understanding (and then mitigating) various procedural biases in S&Is.

**Technical obstacles.** Despite their advancements, AI agents for S&Is face technical challenges, some of which are unique, whereas others are shared with conventional S&I methods. One major issue is the computational cost of deploying these models in real-time applications, which can limit their accessibility, particularly for smaller organizations. Additionally, LLMs can occasionally generate hallucinated or incorrect responses because of their probabilistic nature. Ensuring reliable outputs in high-stakes domains such as healthcare or legal services requires robust screening mechanisms and escalation protocols to handle errors appropriately (Chang et al., 2024). In the future, research will likely focus on improving the efficiency and interpretability of LLMs within AI agents. Reducing computational demands while maintaining high performance is essential for broader adoption. Moreover, ethical considerations around privacy and data security remain critical areas for future development, particularly as LLMs become more integrated into sensitive sectors (Meshkin et al., 2024). While the development and usage of AI agents have become simpler, particularly with the emergence of no-code software development platforms, challenges still exist. Researchers without a technical background may find the process uncomfortable or intimidating, requiring more support and guidance to effectively deploy these systems. Furthermore, domain expertise plays a crucial role in ensuring that the knowledge base of these AI agents is accurate, comprehensive, and tailored to the specific needs of the research. The proposed modular approach in this study addresses these concerns by explicitly considering domain-specific knowledge in the modular structure, allowing experts to contribute to the knowledge base through structured and customizable inputs. This feature ensures that specialized knowledge bases, including documents, dictionaries, handbooks, wikis, and knowledge graphs, can be effectively utilized to support S&Is driven by AI agents. In relation to the bottleneck of concept and relationship extraction, we recognize the challenges posed by polysemy (one word or phrase referring to multiple concepts) and synonymy (multiple words or phrases referring to the same concept) (Renu and Mocko, 2016). Additionally, relationship extraction involves complex tasks such as identifying co-occurrence, distinguishing between correlational and causal relationships, and differentiating between positive, negative, parallel, and hierarchical relationships. While addressing these challenges is time-consuming and resource-intensive, they are critical for improving the accuracy and relevance of LLM outputs. The modular approach can help allow domain experts refine and add specific rules or guidelines for concept and relationship extraction, ensuring that the chatbot is better equipped to handle the intricacies of domain-specific knowledge. This enables the system to improve its ability to handle ambiguities and relationships, leading to more accurate and contextually relevant responses.

**Methodological challenges.** S&Is in transportation research also face several methodological challenges. One significant issue is sampling bias, which can occur when the participants surveyed do not accurately represent the broader population (Meyer de Freitas et al., 2019). This challenge is particularly pronounced in transportation studies, where underrepresentation of certain groups, such as non-drivers, individuals in rural areas, or people with sensory, cognitive, and physiological disabilities, can skew results, limit the generalizability of findings, and cause unintended consequences. Moreover, stated preferences, opinions, and expert insights gathered through S&Is are often subject to framing effects, where the format, wording, or structure of questions can unintentionally influence respondents' answers. Even subtle variations in how options are presented may shift preferences or opinions, which can lead to biased results if not carefully managed (Giergiczny et al., 2017; Yu and Jayakrishnan, 2018). Thus, controlling for framing effects is essential to ensure that the data accurately reflect respondents' true preferences and judgments. At the same time, we do want to clarify that the proposed modular approach does not directly aim to reduce framing effects. Instead, it seeks to establish a framework that facilitates the identification and examination of factors contributing to framing effects in AI-aided S&Is, ultimately supporting more transparent and reliable data collection practices in transportation research. Social desirability bias (van de Mortel, 2008) also persists and is influenced by how respondents perceive AI-driven systems. In addition to social desirability bias, experimenter effects such as placebo effects, expectancy effects, and cultural bias may impact respondents' answers, particularly in human-led S&Is. These effects are unclear in chatbot-based surveys—on the one hand, the algorithmic nature of the system reduces human influence on respondents' behavior; on the other hand, the algorithmic nature might introduce new types of effects. Regardless, S&Is led by AI agents cannot eliminate these biases, and mitigating them may require careful system design, such as customizing prompts to control for biases across different contexts. However, whether eliminating these experimenter effects is desirable depends on the specific research goals, as some effects may be beneficial for understanding human perceptions of transportation systems. The stochastic nature of LLMs introduces variability in responses, which can be mitigated by adjusting parameters such as "temperature" to balance consistency and creativity. Sensitivity analyses are needed to optimize this trade-off, ensuring the system's robustness in diverse scenarios. When S&I is conducted in person, experimenter effects, such as unintended influences on participant responses, can be mitigated by using modular AI agents with varied or fine-tuned prompts and UI designs. Aversion to algorithmic agents, however, remains a challenge. Research suggests that some participants may exhibit algorithm aversion, especially when they perceive AI-driven systems as less trustworthy or less capable than human facilitators do (Mahmud et al., 2022). To reduce this aversion, developers can leverage engineered prompts and modules that enhance the agent's transparency, explainability, and rapport-building capabilities. Additionally, ensuring that the system consistently provides relevant, accurate, and contextually appropriate responses may alleviate concerns about the technology's competence. Developers can also address participant concerns through adaptive questioning strategies, thereby ensuring engagement while reducing the likelihood of negative responses due to algorithmic distrust. Finally, framing effects—where question phrasing impacts responses—remain a challenge. Future ongoing assessment is needed to understand their effectiveness, compared with human-led S&I or non-interactive S&Is, in various contexts. Another critical challenge is the discrepancy between stated and revealed preferences, which is partially caused by the framing effects during S&Is (Yu and Jayakrishnan, 2018) and affects both traditional and chatbot-based surveys. Respondent fatigue, a common issue in lengthy surveys, is mitigated through adaptive questioning strategies that maintain engagement. However, monitoring dropout rates remains essential to ensure data quality. AI agents can integrate real-time behavioral data facilitated by biometric sensors (e.g., eye-tracking devices and heart rate variability measures) to bridge this gap, dynamically adjusting questions based on user behavior. Striking the right balance between structured procedures and flexibility in question delivery is crucial to maintaining both the integrity and adaptability of the survey, ensuring a comprehensive understanding of participant preferences and experiences.

**Ethical concerns and implications.** Ethical concerns in transportation research involving the use of AI agents are multi-faceted and require careful consideration. Future research should explore the impact of agent design on biases, particularly the role of anthropomorphism (Adam et al., 2021), as AI's increasing human-like abilities can shape user perceptions and influence behavior. The potential for chatbots to be perceived as human surrogates may foster social relationships and emotional connections between users and AI, which raises questions about how these interactions affect user trust and decision-making and even lead to AI obsession. It is essential to investigate how chatbots' behavioral characteristics and their perceived social roles influence user engagement, particularly in vulnerable populations. Another key challenge is privacy concerns, particularly in both small-sample studies and large-scale public surveys (Ge and Fukuda, 2016). Anonymizing respondents is vital to protect individual privacy, especially in areas with small populations where identifying respondents from the data is easier. To mitigate this, researchers employ rigorous anonymization techniques to ensure that data cannot be traced back to individuals while maintaining the richness of the data for analysis. The proposed modular approach further addresses this concern by differentiating on-premise LLMs and cloud-based (commercialized) LLMs. However, striking the right balance between privacy protection and data granularity remains an ongoing challenge in transportation research. Despite these challenges, S&Is remain indispensable tools in transportation research. The ethical deployment of AI agents in S&Is is crucial, especially given their potential for collecting sensitive data. Several key ethical concerns must be addressed, including biased question framing, unintended emotional impacts on respondents, and ensuring transparency in how AI-generated decisions are made. Although privacy measures, such as anonymization and secure data storage, have been integrated, these systems must continuously evolve to comply with data privacy regulations such as the GDPR and PIPEDA (Ghandour et al., 2024). Ensuring transparency in the decision-making process of AI, especially in terms of how it frames questions or makes follow-up decisions, is also essential for building trust among users. Moreover, the risk of exacerbating digital divides through reliance on participants' personal devices is significant. Those without access to appropriate technology or internet connectivity could be marginalized, potentially skewing data collection and public participation. To address this, organizations conducting these surveys should consider providing devices or access to digital tools and internet-compatible services (e.g., a "Lite version" of the agent), particularly in underserved communities. This step is necessary for ensuring inclusivity and broadening participation in chatbot-based S&Is. Beyond accessibility, it is also critical to consider the emotional and psychological impacts of interacting with AI-driven systems. As AI agents become more sophisticated, they might unintentionally evoke strong emotional responses or influence decision-making through subtle nudges. Safeguarding ethical standards around these interactions, including exploring how much transparency is needed for users to feel comfortable with AI, should be a priority. Furthermore, while it may seem advantageous to present chatbots as human-like agents to encourage engagement, this approach could inadvertently lead to over-trust and unrealistic expectations of AI capabilities. This is especially concerning for vulnerable populations such as elderly individuals and adolescents, who may be more susceptible to scams or privacy violations. Transparent communication about the limitations and capabilities of the AI system is therefore critical to maintaining ethical standards and protecting participants' privacy. Addressing these challenges is essential to ensure that AI-driven S&Is are both effective and ethically sound.

The modular framework presented in this paper contributes directly to the realization of Human-Machine Transportation Systems (HMTS), where humans and machines collaboratively drive transportation system planning, implementation, operation, and usage (Yu, 2018; Yu and Hyland, 2025). The AI agents (including those with robotic capabilities such as autonomous vehicles) support HMTS by enabling real-time data collection and facilitating collaborative decision-making between human users and AI systems. Although the current implementation focuses on S&Is, there is significant potential to expand AI agents to a wider range of applications, such as focus groups, Delphi methods, and participatory planning processes. That is, AI agents can role play various system stakeholders and communicate with (conducting S&Is on) each other. In the particular context of AI-facilitated participatory planning, agents can facilitate iterative feedback loops and group consensus building, making these processes more scalable and inclusive, offering new opportunities for engaging diverse community stakeholders and obtaining a broad range of perspectives (Yu and McKinley, 2024). Additionally, AI agents have also become important interface components in traveler information applications and autonomous vehicles. In the future, incorporating insights from psychology, behavioral economics, human-computer interaction (HCI), and ethics will be crucial for creating agents that not only collect data efficiently but also engage respondents in culturally sensitive and empathetic ways.

## 6. Conclusions

In this paper, we introduce a novel modular framework for AI agents tailored to conducting S&Is in transportation research. By integrating engineered prompts, customizable knowledge bases, and adaptive conversational logic, the proposed system addresses key limitations of traditional S&Is, such as interviewer bias, scalability, privacy preservation, token use efficiency, and response accuracy. While the framework itself does not eliminate biases inherent in S&Is, its parameterized description of the information process facilitates the explicit identification and examination of underlying assumptions, enhancing transparency in the agent’s actions. Through three field studies—eliciting resident travel preferences under different weather conditions, public perceptions of a new infrastructure project, and expert opinions on the impact of novel technologies—we demonstrate the versatility and effectiveness of the modular framework in a range of contexts. These experiments highlight the resultant agent’s ability to provide real-time clarifications, handle multilingual and multimodal inputs, and ensure robust data collection across diverse participant groups. The proposed AI approach offers a comprehensive solution to various types of S&Is in transportation research, contributing to the next stage for future data collection in the field.

Despite these advancements, our findings also highlighted several challenges in developing and deploying AI agents, particularly concerning framing effects, social desirability biases, and the accessibility of the AI agents (when used for public consultation and participatory decision-making). All methods except the last one are shared by other existing S&I methods. Future improvements must focus on minimizing these biases while enhancing the system’s capacity for handling more complex and dynamic interactions. As we observed in our studies, the stochastic nature of LLMs and the need for ongoing sensitivity analyses to fine-tune agent behavior are critical areas for further research, especially in optimizing the balance between response creativity and consistency. In terms of the immediate next steps, the development of AI agents for S&I will focus on refining adaptability and ensuring robustness in real-world applications. This includes enhancing the agent's ability to tailor follow-up questions based on individual responses, which can improve data accuracy and participant engagement. Expanding the framework's multilingual capabilities and incorporating more advanced voice recognition features will make the system even more accessible, especially in diverse, large-scale studies. Further exploration of reducing token consumption and streamlining data processing will also be crucial for optimizing the agent's performance cost-efficiently, especially for longer S&Is involving complex topics. Additionally, future research will involve applying these agents across an even wider range of transportation research contexts, such as traveler information provisions, public transit user feedback, and consumer support (e.g., purchasing electric and autonomous vehicles). We hope that the proposed framework can lead to more informed and timely transportation decisions and enable practitioners and researchers to engage with larger, more diverse populations without sacrificing the depth or quality of insights.

## Replication and data sharing

The data is available upon request.

## Acknowledgements

We extend our sincere appreciation to all the participants who contributed to the testing and deployment phases of the surveys and interviews, both in person and virtually. While we are unable to mention individual names owing to strict anonymity requirements, we deeply value the insights and contributions from each participant.

**For Experiment 1**, we are thankful for the financial support provided by Environment and Climate Change Canada (ECCC) and the support from Narges Ahmadi (Ph.D. candidate in civil engineering at McGill University) and Nelly Nicola (M.S. student in civil engineering at McGill University). **For Experiment 2**, we greatly appreciate the City of Candiac's support, along with contributions from Ana María Ospina Salazar (M.S. candidate in civil engineering at McGill University), Nicolás Alessandroni (FRQSC postdoctoral fellow in psychology at Concordia University), and Alejandro Pérez Villaseñor (Ph.D. candidate in civil engineering at McGill University). **For Experiment 3**, we are grateful for the support from the Mens, Manus, and Machina (M3S), as part of the Massachusetts Institute of Technology (MIT)'s SMART Program in Singapore. We also appreciate the staff of MIT Mobility Initiatives (MMI), especially Bhuvan Atluri, and the valuable feedback from students and researchers at the MIT JTL Urban Mobility Lab, including Raha Peyravi, Amelia Baum, Awad Abdelhaim, and Hanyong Tang. We also acknowledge the input from Ph.D. candidate Fuqiang Liu at McGill University, who provided high-quality responses and some design suggestions in the early stage of agent development. **For all three experiments**, the presented results are preliminary for the purpose of demonstrating the modular framework and not final. Any mistakes are the sole responsibility of the authors.

Finally, we would like to thank Prof. Michel Belier from the University of Sydney for offering insightful feedback and suggestions.

## Declaration of competing interest

The authors declare that they have no known competing financial interests or personal relationships that could have appeared to influence the work reported in this paper.

## Appendix

## Example dialog from the semi-structured interview in Experiment 3

To better demonstrate the ability of an AI agent to conduct semi-structured interview based on the proposed system architecture and procedure, we present an example. Please note that the transcripts are verbatim. Therefore, any typos are also kept. The text shifted to the left side is from the agent, whereas the text shifted to the right side is from the participant.

Hello! My name is RoBot. I want to learn your thoughts about the potential adoption and impact of Large Language Models (LLMs) in transportation and urban mobility.

> Understood (Button)

This interview comprises only five questions, so I encourage you to share your detailed thoughts and perspectives based on your expertise and experience.
As a respected professional in this field, your participation is not only highly appreciated but also pivotal in driving the human-centered advancement of AI technologies within the transportation sector.

> Sounds good (Button)

**Question 1/5:** Could you describe any specific applications of LLMs in your work, organization, or field?

> I mostly use it to learn a new thing in my field (such as a method in Transportation Engineering or the usage of different ML algorithms on databases). For example, I use chatGPT to learn more about the performance of XGBoost in the classification of imbalanced datasets and how to find the best hyperparameters for my model. Another use case is to summarize papers so I can read them in less time and ask it to explain anything I am not familiar with inside the paper.

Great! So you have mentioned that you primarily utilize the tool to acquire new knowledge within your area, such as Transportation Engineering methods or applying various machine learning algorithms to datasets. Specifically, you have used ChatGPT to gain insights into the efficacy of XGBoost for sorting unbalanced datasets and to determine the optimal hyperparameters for your models. Additionally, you employ it to condense research papers, allowing for quicker review, and to clarify any unfamiliar concepts within those papers. This is quite interesting. Thank you for your response.

> Go to the next question (Button)

**Question 2/5:** What are the primary benefits and challenges associated with the use of LLMs in transportation, either from your own experience or based on what you are aware of?"

> Regarding the benefits, I would like to mention its power to brainstorm ideas to implement on your projects and teach you any concept that don't know. Additionally, because most students of Transportation Engineering are now familiar with ML and its usage in this field, it can also help you to write codes and have a powerful tool to implement your analysis on transportation datasets. Also, you can see if you can use an idea from an irrelevant field to use it in Transportation Engineering. I have nothing regarding challenges in mind right now.

I understand that you have highlighted several benefits of large language models, such as their ability to brainstorm ideas, teach concepts, assist with code writing, and transfer ideas from other fields to transportation engineering. However, could you please provide more details on how exactly these models have been applied in your projects or in the classroom? Additionally, while you mentioned not having challenges in mind right now, could you perhaps think of any potential limitations or areas where LLMs might fall short in the context of transportation research?

To provide you with my experience in using LLMs for my projects, I used it to find an open-source dataset related to Transportation so I can implement ML algorithms on them. It found a dataset, suggested a project that I can do with that data, and helped me to correct my code to build a prediction model based on them. I also learned about a new library to find about the importance of my input features of my model as well. Regarding limitations, you usually find wrong citations and fake links when you ask them to provide you with papers related to a field, especially when you say I want the most recent papers related to a subject.

I see, so you have utilized LLMs in your projects, specifically for identifying an open-source transportation dataset to apply machine learning algorithms. The LLM not only located a dataset but also recommended a potential project, assisted in debugging your code, and introduced you to a new library to assess the significance of your model's input features. However, you have encountered limitations, particularly with LLMs providing incorrect citations and unreliable links when requesting recent papers in your field. This is interesting. Thank you for sharing.

Go to the next question (Button)

**Question 3/5:** How do you foresee the applications and impacts of LLMs evolving in both the near term (1–2 years) and the long term (5–10 years) in transportation?

In short-term, we can fill the void of programming that most people in transportation are not professional in. Also, it helps us to make the process of learning in this field much more faster, so we will probably witness more number of published papers in this field in coming years. In long-term, we will see that people in this field will use more advanced technology and techniques to develope new methods to control and manage transportation, mostly because we won't lye on universities to learn the knowledge of transportation and we will be able to have the ability of coding much more easier in the next 5-10 years and maybe we won't need to code anymore and LLMs will do the tasks with high performance. Also, I believe we will make changes in the process of writing papers in the nex 5-10 years and will allow LLMs to help us write the most of the papers and we will push the boundaries of science in transportation much more faster.

I see. So you mentioned that in the short term, LLMs will bridge the gap in programming expertise among transportation professionals, accelerating the learning curve and potentially increasing the volume of research publications in the field. Looking ahead, you anticipate that within the next 5–10 years, transportation experts will increasingly adopt sophisticated technology and methodologies for traffic management, partly because reliance on traditional university education for transportation knowledge will diminish. You also foresee a future where coding becomes much simpler or perhaps even unnecessary, as LLMs may take over such tasks with high efficiency. Furthermore, you expect that the process of academic writing will evolve, with LLMs playing a significant role in drafting research papers, thereby expediting scientific advancements in transportation. This is helpful. Thank you for sharing.

I want to add or clarify (Button)

Great. I am listening : )

Also, we will experience the advent of AVs more than before. The infrastructure of transportation will become more intelligent with the use of AI and Intelligent Transportation Systems such as adaptive traffic signs, adaptive traffic signals, communication between vehicles and infrastructure to update traffic data on real time so we can manage traffic with the help of AI as well. The future of transportation is bright :)

**Question 4/5:** What do you think are the main barriers (e.g., technological, regulatory, and economic) that currently hinder the broader adoption of LLMs in transportation, and how might these be addressed?

Respectfully, I believe there are a lot of conservative people in this field who are against the new ways of using such tools. For example, it is still forbidden to get help from LLMs to write papers, because academic people can't trust it. However, if we make this process more automatic, we will be able to push the boundaries of science more faster. Also, safety issues in using them for our daily use is considered as one of the barries, such as giving informations to the LLMs might end up to sharing your sensitive information with government. Economicaly, this technology will make too many people unemployed and this can have bad affects on the sociaty and the moral situation of people. It is also expensive to provide the infrastructure of city to adapt it to the new era of AI to make everything automatic and adoptable with LLMs and their usage.

I understand that you have concerns about the conservative stance of some professionals in the field and the potential risks associated with using LLMs, including privacy and employment issues. Could you please provide more details on how you think these barriers could be addressed to facilitate the adoption of large language models in transportation and urban mobility?

Regarding the safety issues, it is important to monitor companies which develope such technologies and make regulations on these types of privacy data gathering issues. But regarding unemployment risk, I think we have no choice but to accept it and try to find new ways of employing people to help transportation and urban mobility. However, regarding the acceptance of LLMs to be involved in writing papers, I belive the best way is to dedicate new journals to them and see how they would perform so we can understand if it is beneficial or not.

I see, so you're emphasizing the need for oversight and regulation in regard to safety and privacy concerns raised by the development of large language models. You also acknowledge the inevitable impact on employment but suggest we should adapt by creating new job opportunities within transportation and urban mobility sectors. Additionally, you propose an intriguing idea of establishing specialized journals for papers written with the help of large language models, to evaluate their effectiveness and contribution to the field. This is interesting. Thank you for sharing.

Go to the next question (Button)

Almost There!

**Question 5/5:** Using your creative imagination, can you envision scenarios where the integration of LLMs might dramatically transform transportation systems or shape policymaking? What are the anticipated benefits and potential challenges of such changes?

1. Traffic Management and Optimization Real-Time Traffic Analysis: LLMs can process vast amounts of data from sensors, cameras, and GPS devices to analyze traffic patterns in real time. They can provide insights and predictions about congestion, accidents, and optimal routes. Incident Management: By analyzing reports and social media posts, LLMs can quickly identify traffic incidents and suggest alternative routes or solutions to minimize delays. 2. Public Transportation Enhancement Dynamic Scheduling: LLMs can optimize bus and train schedules based on real-time demand, reducing wait times and improving service efficiency. Personalized Travel Assistance: Chatbots powered by LLMs can assist passengers with route planning, ticket booking, and real-time updates on delays or changes. 3. Autonomous Vehicles Natural Language Processing: LLMs can enhance the communication systems of autonomous vehicles, enabling them to understand and respond to human commands more effectively. Decision-Making Support: LLMs can analyze complex scenarios and provide decision-making support for autonomous driving systems, improving safety and efficiency. 4. Logistics and Freight Management Route Optimization: By analyzing weather conditions, traffic, and delivery constraints, LLMs can help logistics companies optimize delivery routes, saving time and reducing fuel consumption. Supply Chain Management: LLMs can predict demand fluctuations, manage inventory levels, and streamline supply chain operations, leading to more efficient transportation of goods. 5. Infrastructure Planning and Maintenance Predictive Maintenance: By analyzing data from sensors embedded in infrastructure, LLMs can predict maintenance needs, preventing breakdowns and extending the lifespan of transportation assets. Smart Infrastructure Design: LLMs can assist in designing smarter transportation networks by analyzing urban development patterns and predicting future transportation needs. 6. Sustainability and Environmental Impact Emission Reduction: LLMs can help in planning routes and schedules that minimize fuel consumption and emissions, contributing to greener transportation systems. Policy Development: By analyzing vast datasets, LLMs can support policymakers in developing regulations and initiatives to promote sustainable transportation. 7. User Experience Enhancement Customer Service: LLM-powered virtual assistants can provide 24/7 support to travelers, addressing their concerns and queries in multiple languages. Accessibility: LLMs can improve accessibility by offering tailored information and assistance to people with disabilities, ensuring they can navigate transportation systems more easily. 8. Data-Driven Decision Making Predictive Analytics: LLMs can forecast transportation trends and usage patterns, helping stakeholders make informed decisions about investments and developments. Risk Management: By analyzing historical data and current trends, LLMs can identify potential risks and vulnerabilities in transportation systems, enabling proactive measures. The most important challenge is data privacy, fast reaction to mistakes from LLMs, giant companies that try to prevent expansion usage of these technologies becuase their profit might be in old ways.

Great! So you mentioned that traffic management and optimization can be improved through the use of LLMs. They can analyze real-time traffic data from various sources like sensors, cameras, and GPS devices to provide insights and predictions about congestion, accidents, and optimal routes. Additionally, LLMs can help in incident management by quickly identifying traffic incidents through analysis of reports and social media posts, and suggesting alternative routes or solutions to minimize delays.

You also highlighted that LLMs can enhance public transportation by optimizing bus and train schedules based on real-time demand, reducing wait times and improving service efficiency. Moreover, chatbots powered by LLMs can provide personalized travel assistance to passengers, helping with route planning, ticket booking, and real-time updates on delays or changes.

In the context of autonomous vehicles, LLMs can enhance communication systems by enabling them to understand and respond to human commands more effectively through natural language processing. They can also analyze complex scenarios and provide decision-making support for autonomous driving systems, improving safety and efficiency.

Furthermore, LLMs can assist in logistics and freight management by optimizing delivery routes based on weather conditions, traffic, and delivery constraints, thus saving time and reducing fuel consumption. They can also predict demand fluctuations, manage inventory levels, and streamline supply chain operations, leading to more efficient transportation of goods.

In terms of infrastructure planning and maintenance, LLMs can predict maintenance needs by analyzing data from sensors embedded in infrastructure, preventing breakdowns and extending the lifespan of They also assist in designing smarter networks by analyzing urban development patterns.

For sustainability, LLMs can plan routes to minimize fuel consumption and emissions, supporting greener transportation systems. They aid policymakers in developing sustainable transportation regulations.

LLMs enhance user experience with 24/7 virtual assistants for customer service, providing multilingual support and tailored information for people with disabilities, ensuring accessibility.

In data-driven decision-making, LLMs forecast trends and usage patterns, helping stakeholders make informed decisions. They also identify risks in transportation systems through historical data analysis, enabling proactive measures.

However, challenges include ensuring data privacy, swiftly addressing LLM mistakes, and overcoming resistance from companies invested in traditional methods.

Overall, LLM integration in transportation offers significant benefits but requires careful management of privacy, accuracy, and industry resistance. Thank you for your insights.

Proceed (Button)

Thank you for answering these questions!

## Author biography

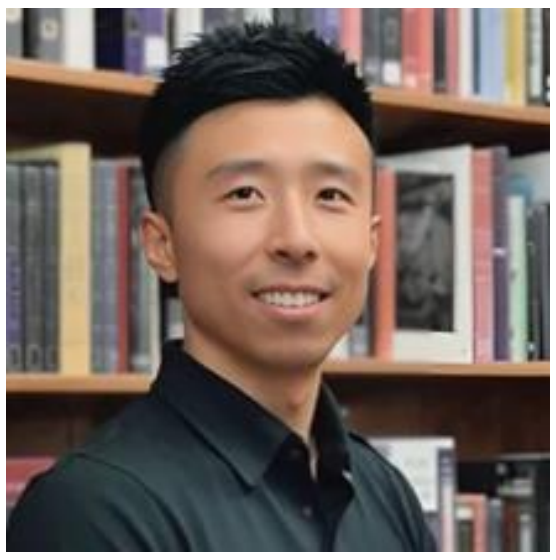

**Jiangbo Yu** is an Assistant Professor in the Department of Civil Engineering at McGill University. He is a registered Professional Engineer (PE) and a licensed Professional Transportation Planner (PTP). He holds a B.Sc. from the School of Mechanical Engineering at Beijing Institute of Technology, an M.Sc. in Civil Engineering from the University of Southern California, and a Ph.D. in Civil Engineering from the University of California, Irvine. His research focuses on human-machine collaborative engineering and planning of transportation systems.

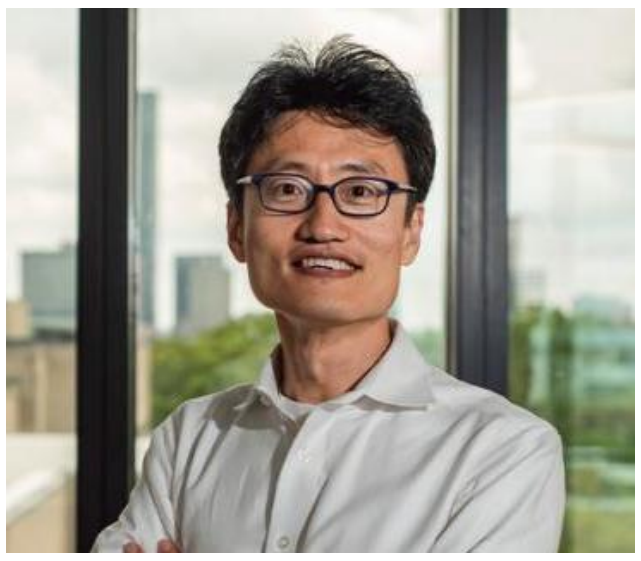

**Jinhua Zhao** is the Professor of Cities and Transportation at the Massachusetts Institute of Technology (MIT). Prof. Zhao integrates behavioral and computational thinking to decarbonize the world's mobility system. Prof. Zhao founded the MIT Mobility Initiative, coalescing the Institute's efforts on transportation research, education, entrepreneurship, and engagement. He hosts the MIT Mobility Forum, highlighting transportation innovation from MIT and across the globe. Prof. Zhao directs the JTL Urban Mobility Lab and Transit Lab, leading long-term collaborations with transportation authorities and operators worldwide and enabling cross-culture learning between cities in North America, Asia and Europe.

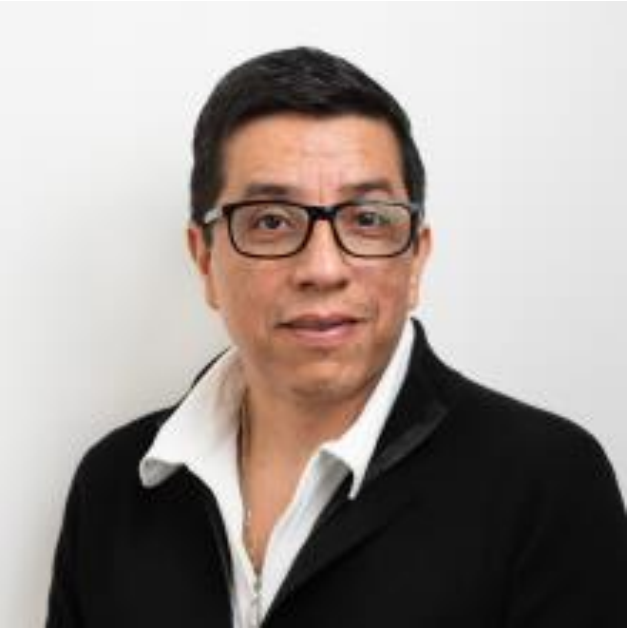

**Luis Miranda-Moreno** is an Associate Professor at the Department of Civil Engineering at McGill University. His scientific and professional contributions have been mostly in the areas of sustainable urban mobility and road safety. His areas of research and expertise include emission modeling methods, the impact assessment of built environment and road design strategies on climate and road injuries, the strategies to promote non-motorized urban mobility and the integration of emerging technological solutions.

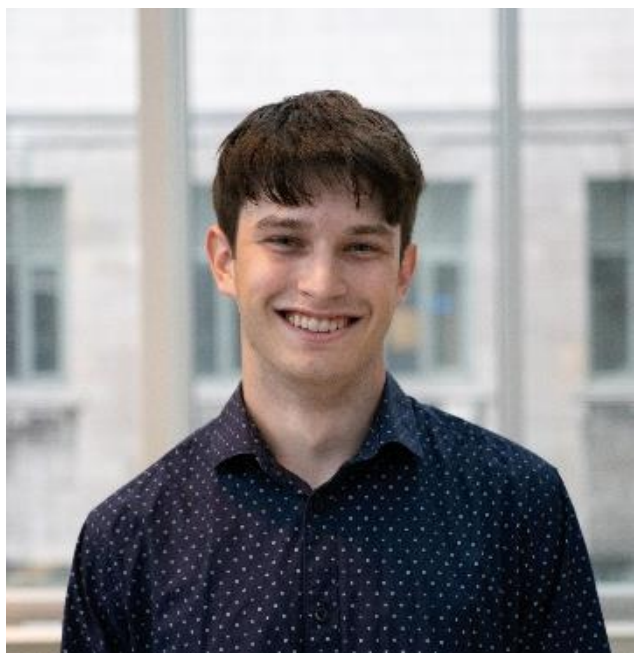

**Matthew Korp** is an undergraduate researcher in the Faculty of Engineering at McGill University. He is actively engaged in Prof. Jiangbo Yu's research on generative AI for urban mobility engineering and planning and serves as the Ejection Co-Lead for the McGill Rocket Team.